%% file: ms.tex
\documentclass[fleqn, usenatbib]{mnras}
\usepackage{style}

\title[Improving Galaxy Clustering with Deep Learning]{Improving Galaxy Clustering Measurements with Deep Learning: analysis of the DECaLS DR7 data}

\author[M. Rezaie et al.]{
Mehdi Rezaie$^{1}$\thanks{E-mail: mr095415@ohio.edu},
Hee-Jong Seo$^{1}$\thanks{E-mail: seoh@ohio.edu},
Ashley J. Ross$^{2}$,
and Razvan C. Bunescu$^{3}$
\\
$^{1}$Department of Physics and Astronomy, Ohio University, Athens, OH 45701, USA\\
$^{2}$The Center of Cosmology and Astro Particle Physics, the Ohio State University, Columbus, OH 43210, USA\\
$^{3}$School of Electrical Engineering and Computer Science, Ohio University, Athens, OH 45701, USA
}

\begin{document}\label{firstpage}
\pagerange{\pageref{firstpage}--\pageref{lastpage}}
\maketitle

\begin{abstract}
Robust measurements of cosmological parameters from galaxy surveys rely on our understanding of systematic effects that impact the observed galaxy density field. In this paper we present, validate, and implement the idea of adopting the systematics mitigation method of Artificial Neural Networks for modeling the relationship between the target galaxy density field and various observational realities including but not limited to Galactic extinction, seeing, and stellar density. Our method by construction allows a wide class of models and alleviates over-training by performing k-fold cross-validation and dimensionality reduction via backward feature elimination. By permuting the choice of the training, validation, and test sets, we construct a selection mask for the entire footprint.  We apply our method on the extended Baryon Oscillation Spectroscopic Survey (eBOSS) Emission Line Galaxies (ELGs) selection from the Dark Energy Camera Legacy Survey (DECaLS) Data Release 7 and show that the spurious large-scale contamination due to imaging systematics can be significantly reduced by up-weighting the observed galaxy density using the selection mask from the neural network and that our method is more effective than the conventional linear and quadratic polynomial functions. We perform extensive analyses on simulated mock datasets with and without systematic effects. Our analyses indicate that our methodology is more robust to overfitting compared to the conventional methods. This method can be utilized in the catalog generation of future spectroscopic galaxy surveys such as eBOSS and Dark Energy Spectroscopic Instrument (DESI) to better mitigate observational systematics.
\end{abstract}



\input{sections/introduction}
\input{sections/data}
\input{sections/methodology}

\input{sections/results}
\input{sections/conclusion}
\input{sections/acknowledgement}

\bibliographystyle{mnras}
\bibliography{refs}

\appendix
\input{sections/window}


\bsp	
\label{lastpage}
\end{document}

%% file: sections/introduction.tex
\section{Introduction}\label{sec:intro}
Our current understanding of the Universe is founded upon statistical analyses of cosmological observables, such as the large-scale structure of galaxies, the cosmic microwave background, and the Hubble diagram of distant type Ia supernovae \citep[e.g.,][]{efstathiou1988analysis, fisher1993power, smoot1992structure, mather1994measurement,riess1998observational, perlmutter1999measurements, ata2017clustering, BOSSfinal, jones2018measuring, akrami2018planck, Elvin18}. Among these probes, galaxy surveys aim at constructing clustering statistics of galaxies with which we can investigate the dynamic of the cosmic expansion due to dark energy, test Einstein's theory of gravity, and constrain the total mass of neutrinos and statistical properties of the primordial fluctuations, etc \citep{peebles1973statistical,kaiser1987clustering,mukhanov1992theory,hamilton1998linear,eisenstein1998cosmic,seo2003probing,eisenstein2005dark,sanchez2008best, dalal2008imprints}.\\

The field of cosmology has been substantially advanced by the recent torrents of spectroscopic and imaging datasets from galaxy surveys such as the Sloan Digital Sky Survey (SDSS), Two Degree Field Galaxy Redshift Survey, and WiggleZ Dark Energy Survey \citep{york2000sloan, colless20012df, drinkwater2010wigglez}. The SDSS has been gathering data through different phases SDSS-I (2000-2005), SDSS-II (2005-2008), SDSS-III (2008-2014), and SDSS-IV (2014-2019). In order to derive more robust constraints with higher statistical confidence along with advancements in the technology of spectrographs, software, and computing machines, future large galaxy surveys aim at not only a wider area but also fainter galaxies out to higher redshifts. As an upcoming ground-based survey, the Dark Energy Spectroscopic Instrument (DESI) experiment will gather spectra of thirty million galaxies over 14,000 deg$^{2}$ starting in late 2019; this is approximately a factor of ten increase in the number of galaxy spectra compared to those observed in SDSS I--IV. This massive amount of spectroscopic data will lead to groundbreaking measurements of cosmological parameters through statistical data analyses of the clustering measurements of the 3D distribution of galaxies and quasars \citep{aghamousa2016desi}.\\

The Large Synoptic Survey Telescope (LSST) is another ground-based survey currently being constructed. It will gather 20 Terabytes of imaging data every night and will cover 18,000 deg$^{2}$ of the sky for ten years with a sample size of $2\times 10^{9}$ galaxies. The LSST would provide enough imaging data to address many puzzling astrophysical problems from the nature of dark matter, the growth of structure to our own Milky Way. Given such a data volume, many anticipate that the LSST will revolutionize the way astronomers do research and data analysis \citep{ivezic2008lsst, LSSTObservingStrategyWhitePaper}. \\

The enormous increased data volume provided by DESI and the LSST will significantly improve statistical confidences but will require analyses that are more complex and sensitive to the unknown systematic effects. A particular area of concern is the systematic effects due to imaging attributes such as atmospheric conditions, foreground stellar density, and/or inaccurate calibrations of magnitudes. These systematic effects can affect the target galaxy selections and therefore induce the non-cosmological perturbations into the galaxy density field, leading to excess clustering amplitudes, especially on large scales \citep[see e.g.][]{myers2007clustering,thomas2011angular,thomas2011excess, ross2011ameliorating, ashley2012MNRAS, 2012ApJ...761...14H, huterer2013calibration, pullen2013systematic}.\\

Robust and precise measurements of cosmological parameters from the large-scale galaxy clustering are contingent upon thorough treatment of such systematic effects. Many techniques have been developed to mitigate the effects. One can generally classify these methods into the mode projection, regression,  and Monte Carlo simulation of fake objects.\\ 

The mode-projection based techniques attribute a large variance to the spatial modes that strongly correlate with the potential systematic maps such as imaging attributes, thereby effectively removing those modes from the estimation of power spectrum \citep[see e.g.][]{rybicki1992interpolation,tegmark1997measure,tegmark1998measuring,slosar2004exact,ho2008correlation, pullen2013systematic,leistedt2013estimating,leistedt2014exploiting}. In detail, the basic mode projection \citep{leistedt2013estimating} produces an unbiased power spectrum and is equivalent to a marginalization over a free amplitude for the contamination produced by a given map.
 The caveat is that the variance of the estimated clustering increases by projecting out more modes for more imaging attributes. The \textit{extended} mode projection technique \citep{leistedt2014exploiting} resolves this issue by selecting a subset of the imaging maps using a $\chi^{2}$ threshold to determine the significance of a potential map. The limitation of the mode-projection based methods is that they are only applicable to the two-dimensional clustering measurements, and they reintroduce a small bias \citep{elsner2015unbiased}. \citet{kalus2016unbiased} extended the idea to the 3D clustering statistics and developed a new step to unbias the measurements~\citep[for an application on SDSS-III BOSS data see e.g.,][]{kalus2018map}\\

The regression-based techniques model the dependency of the galaxy density on the potential systematic fluctuations and estimate the parameters of the proposed function by solving a least-squares problem, or by cross-correlating the galaxy density map with the potential systematic maps \citep[see e.g.][]{ross2011ameliorating, ashley2012MNRAS,Ross17,2012ApJ...761...14H,delubac2016sdss, prakash2016sdss, Raichoor2017MNRAS.471.3955R, laurent2017clustering, Elvin18, 2018ApJ...863..110B}. The best fit model produces a \textit{selection mask (function)} or a set of \textit{photometric weights} that quantifies the systematic effects in the galaxy density fluctuation induced by the imaging pipeline, survey depth, and other observational attributes. The selection mask is then used to up-weight the observed galaxy density map to mitigate the systematic effects. The regression-based methods often assume a linear model (with linear or quadratic polynomial terms), and use all of the data to estimate the parameters of the given regression model; however, the assumption that the systematic effects are linear might not necessarily hold for strong contamination, e.g., close to the Galactic plane. \citet{2012ApJ...761...14H} analyzed photometric Luminous Red Galaxies in SDSS-III Data Release 8 and showed that the excess clustering due to the stellar contamination on large scales (e.g., roughly greater than twenty degrees) cannot be removed with a linear approximation. \citet{rossfNL} investigated the local non-Gaussianity ($f_{NL}$) using the BOSS Data Release 9 ``CMASS" sample of galaxies \citep{ahn2012ninth} and found that a robust cosmological measurement on very large scales is essentially limited by the systematic effects. Their analysis indicated that a more effective systematics correction is preferred relative to the selection mask based on the linear modeling of the stellar density contamination. Recently, \citet{Elvin18} developed a methodology based on $\chi^{2}$ statistics to rank the imaging maps based on their significance, and derived the selection mask by regressing against the significant maps.\\

Another promising, yet computationally expensive, approach injects artificial sources into real imaging in order to forward-model the galaxy survey selection mask introduced by real imaging systematics \citep[see e.g.][]{berge2013ultra,Balrog}. Rapid developments of multi-core processors and efficient compilers will pave the path for the application of these methods on big galaxy surveys.\\

In this paper, we develop a systematics mitigation method based on artificial neural networks. Our methodology models the galaxy density dependence on observational imaging attributes to construct the selection mask, without making any prior assumption of the linearity of the fitting model. Most importantly, this methodology is less prone to over-training and the resulting removal of the clustering signal by performing $k$-fold cross-validation (i.e., splitting the data into $k$ number of groups/partitions from which one constructs the training, validation, and the test sets) and dimensionality reduction through backward feature elimination~(i.e., removing redundant and irrelevant imaging attributes)\citep[see e.g.,][]{devijver1982pattern, john1994irrelevant, koller1996toward, kohavi1997wrappers, ramaswamy2001multiclass, guyon2003introduction}.  By permutation of the training, validation, and test sets, the selection mask for the entire footprint is constructed.  We apply our method on galaxies in the Legacy Surveys Data Release 7 (DR7) \citep{dey2018overview} that are chosen with the eBOSS-ELG color-magnitude selection criteria~\citep{Raichoor2017MNRAS.471.3955R} and compare its performance with that of the conventional, linear and quadratic polynomial regression methods. While the effect of mitigation on the data will be estimated qualitatively as well as quantitatively based on cross-correlating the observed galaxy density field and imaging maps, the data does not allow an absolute comparison to the unknown underlying cosmology. We therefore simulate two sets of 100 mock datasets, without and with the systematic effects that mimic those of DR7, apply various mitigation techniques in the same way we treat the real data, and test the resulting clustering signals against the ground truth.\\

This paper is organized as follows. Section \ref{sec:data} presents the imaging dataset from the Legacy Surveys DR7 used for our analysis. In Section \ref{sec:method}, we describe our method of Artificial Feed Forward Neural Network in detail as well as the conventional multivariate regression approaches. In this section, we also explain the angular clustering statistics employed to assess the level of systematic effects and the mitigation efficiency. We further describe the procedure of producing the survey mocks with and without simulated contaminations. In Section \ref{sec:results}, we present the results of mitigation for both DR7 and the mocks. Finally, we conclude with a summary of our findings and a discussion of the benefits of our methodology for future galaxy surveys in Section \ref{sec:conclusion}.\\

%% file: sections/data.tex
\section{Legacy Surveys DR7}\label{sec:data}
We use the seventh release of data from the Legacy Surveys \citep{dey2018overview}. The Legacy Surveys are a group of imaging surveys in three optical (r, g, z) and four Wide-field Infrared Survey Explorer (W1, W2, W3, W4; \citet{wright2010wide}) passbands that will provide an inference model catalog amassing 14,000 deg$^{2}$ of the sky in order to pre-select the targets for the DESI survey \citep{lang2016tractor, aghamousa2016desi}. Identification and mitigation of the systematic effects in the selection of galaxy samples from this imaging dataset are of vital importance to DESI, as spurious fluctuations in the target density will likely present as fluctuations in the transverse modes of the 3D field and/or changes in the shape of the redshift distribution. Both effects will need to be modeled in order to isolate the cosmological clustering of DESI galaxies. The ground-based surveys that probe the sky in the optical bands are the Beijing-Arizona Sky Survey (BASS) \citep{zou2017project}, DECam Legacy Survey (DECaLS) and Mayall z-band Legacy Survey (MzLS)\citep[see e.g.,][]{dey2018overview}. Additionally, the Legacy Surveys program takes advantage of another imaging survey, the Dark Energy Survey, for about 1,130 deg$^{2}$ of their southern sky footprint \citep{dark2005dark}. DR7 is data only from DECaLS, and we refer to this data interchangeably as DECaLS DR7 or DR7 hereafter.\\

We construct the ELG catalog by adopting the Northern Galactic Cap eBOSS ELG color-magnitude selection criteria from \citet{Raichoor2017MNRAS.471.3955R} on the DR7 sweep files \citep{dey2018overview} with a few differences in the clean photometry criteria (see Table \ref{tab:ts}). In detail, the original eBOSS ELG selection is based on DR3 while ours is based on DR7. Since the data structure changed from DR3 to DR7, we use \texttt{brightstarinblob} instead of \texttt{tycho2inblob} to eliminate objects that are near bright stars. In contrast to the original selection criteria, we do not apply the \texttt{decam\_anymask[grz]=0} cut, as any effect from this cut will be encapsulated by the imaging attributes used in this analysis. Also, we drop the SDSS bright star mask from the criteria, as this mask is essentially replaced by the \texttt{brightstarinblob} mask. After constructing the galaxy catalog, we pixelize the galaxies into a HEALPix map \citep{gorski2005healpix} with the resolution of 13.7 arcmin ($N_{{\rm side}} = 256$)  in \textit{ring} ordering format to create the observed galaxy density map.\\

\begin{table}
  \begin{center}
    \caption{The Northern Galactic Cap color-magnitude selection of the eBOSS Emission Line Galaxies \citep{Raichoor2017MNRAS.471.3955R}. We enforce the same selection for the entire sky. Note that our selection is slightly different from \citet{Raichoor2017MNRAS.471.3955R} in the clean photometry criteria as explained in the main text.}
    \label{tab:ts}
    \begin{tabular}{l|r}
    \hline
    \hline
      \textbf{Criterion} & \textbf{eBOSS ELG}\\
      \hline
      \multirow{3}{*}{\scriptsize{Clean Photometry}} & \scriptsize{0 mag $<$ V $<$ 11.5 mag Tycho2 stars mask}\\
        & \scriptsize{\texttt{BRICK\_PRIMARY}==True}\\
        & \scriptsize{\texttt{brightstarinblob}==False} \\
     \hline
      \scriptsize{[OII] emitters} &  \scriptsize{21.825 $<$ g $<$ 22.9} \\
      \hline 
      \multirow{2}{*}{\scriptsize{Redshift range}} & \scriptsize{-0.068(r-z) + 0.457 $<$ g-r $<$ 0.112 (r-z) + 0.773}\\
 & \scriptsize{0.637(g-r) + 0.399 $<$ r-z $<$ -0.555 (g-r) + 1.901}\\
      \end{tabular}
  \end{center}
\end{table}

We consider a total of 18 imaging attributes as potential sources of the systematic error since each of these attributes can affect the completeness and purity with which galaxies can be detected in the imaging data. We produce the HEALPix maps \citep{gorski2005healpix} with $N_{{\rm side}}=256$ and oversampling of four\footnote{In this context, `oversampling' means dividing a pixel into sub-pixels in order to derive the given pixelized quantity more accurately. For example, oversampling of four means subdividing each pixel into $4^2$ sub-pixels. If the target resolution is $N_{{\rm side}}=256$, the attributes will be derived based on a map with the resolution of 4$\times$256 when oversampling is four.} for these attributes based on the DR7 ccds-annotated file using the \texttt{validationtests} pipeline\footnote{\url{https://github.com/legacysurvey/legacypipe/tree/master/validationtests}} and the code that uses the methods described in \citet{LeistedtMap}. These include three maps of Galactic structure: Galactic extinction \citep{schlegel1998maps}, stellar density from Gaia DR2 \citep{brown2018gaia}, and Galactic neutral atomic hydrogen (HI) column density \citep{bekhti2016hi4pi}. We further pixelize quantities associated with the Legacy Surveys observations, including the total depth, mean seeing, mean sky brightness, minimum modified Julian date, and total exposure time in three passbands (r, g, and z).  For clarity, we list each attribute below:\\

\begin{itemize}
    \item \textbf{Galactic extinction} (\textit{EBV}), measured in magnitudes, is the infrared radiation of the dust particles in the Milky Way. We use the SFD map \citep{schlegel1998maps} as the estimator of the E(B-V) reddening. The reddening is the process in which the dust particles in the Galactic plane absorb and scatter the optical light in the infrared. This reddening effect affects the measured brightness of the objects, i.e., the detectability of the targets. We correct the magnitudes of the objects for the Milky Way extinction prior to the galaxy (\textit{target}) selection using the extinction coefficients of 2.165, 3.214, and 1.211 respectively for r, g, and z bands based on \citet{schlafly2011measuring}.\\
    
    \item \textbf{Galaxy depth} (\textit{depth}) defines the brightness of the faintest detectable galaxy at $5-\sigma$ confidence, measured in AB magnitudes. The measured depth in the catalogs does not include the effect of Galactic extinction (described above), so we apply the extinction corrections to the depth maps in the same manner.\\
    
    \item {\bf Stellar density} (\textit{nstar}), measured in deg$^{-2}$, is constructed by pixelization of the Gaia DR2 star catalog \citep{brown2018gaia} with the g-magnitude cut of 12 < gmag < 17. The stellar foreground affects the galaxy density in two ways. First, the colors of stars overlap with those of galaxies, and consequently stars can be mis-identified as galaxies and included in the sample, which will result in a positive correlation between the stellar and galaxy distribution. Second, the foreground light from stars impacts the ability to detect the galaxies that are behind them, e.g., by directly obscuring their light or by altering the sky background, which will cause a negative correlation between the two distributions. The second effect may reduce the completeness with which galaxies are selected and was the dominant systematic effect on the BOSS galaxies \citep{ashley2012MNRAS}. The Gaia-based stellar map is a biased set of the underlying stars that actually impact the data. Assuming that there exists a non-linear mapping between the Gaia stellar map and the truth stellar population, linear models might be insufficient to fully describe the stellar contamination. This motivates the application of non-linear models.\\
    
    \item \textbf{Hydrogen atom column density} (\textit{HI}), measured in cm$^{-2}$, is another useful tracer of the Galactic structure, which increases at regions closer to the Milky Way plane. The hydrogen column density map is based on data from the Effelsberg-Bonn HI Survey (EBHIS) and the third revision of the Galactic All-Sky Survey (GASS). EBHIS and GASS have identical angular resolution and sensitivity, and provide a full-sky map of the neutral hydrogen column density \citep{bekhti2016hi4pi}. This map provides complementary information to the Galactic extinction and stellar density maps. Hereafter, \textit{lnHI} refers to the natural logarithm of the HI column density.\\

  {\bf Sky brightness} (\textit{skymag}) relates to the background level that is estimated and subtracted from the images as part of the photometric processing. It thus alters the depth of the imaging. It is measured in AB mag/arcsec$^{2}$. \\

    \item {\bf Seeing} (\textit{seeing}) is the full width at half maximum of the point spread function (PSF), i.e., the sharpness of a telescope image, measured in arcseconds. It quantifies the turbulence in the atmosphere at the time of the observation and is sensitive to the optical system of the telescope, e.g., whether or not it is out of focus. Bad seeing conditions can make stars that are point sources appear as extended objects, therefore falsely being selected as galaxies. The seeing in the catalogs is measured in CCD `pixel'. We use a multiplicative factor of 0.262 to transform the seeing unit to arcseconds.\\

    \item {\bf Modified Julian Date} (\textit{MJD}) is the traditional dating method used by astronomers, measured in days. If a portion of data taken during a specific period is affected by observational conditions during that period, regressing against MJD could mitigate that effect.\\

    \item {\bf Exposure time} (\textit{exptime}) is the length of time, measured in seconds, during which the CCD was exposed to the object light. Longer exposures are needed to observe fainter objects. The Legacy Surveys data is built up from many overlapping images, and we map the total exposure time, per band, in any given area. A longer exposure time thus corresponds to a greater depth, all else being equal.
\end{itemize}

As part of the process of producing the maps, we determine the fractional CCD coverage per passband, fracgood ($f_{{\rm pix}}$), within each pixel with oversampling of four. We define the minimum of $f_{{\rm pix}}$ in r, g, and z passbands as the \textit{completeness} weight of each pixel,
\begin{equation}
    \label{eq:comp}
    \text{completeness}~f_{\rm pix} = \min(f_{\rm pix,r}, f_{\rm pix,g}, f_{\rm pix,z}).
\end{equation}

We apply the following arbitrary cuts, somewhat motivated by the eBOSS target selection, on the depth and $f_{{\rm pix}}$ values to eliminate the regions with shallow depth and low pixel completeness due to insufficient available information: \\
\begin{align}\label{eq:depth_cuts}
depth_{r} &\geq 22.0, \\
depth_{g} &\geq 21.4, \nonumber\\
depth_{z} &\geq 20.5, \nonumber\\
{\rm and}~f_{\rm pix}  &\geq 0.2, \nonumber
\end{align}
which results in 187,257 pixels and an effective total area of 9,459 $\deg^2$ after taking $f_{{\rm pix}}$ into account. We report the mean, 15.9-, and 84.1-th percentiles of the imaging attributes on the masked footprint in Tab. \ref{tab:meanstats}.\\

\begin{figure*}
    \centering
    \includegraphics[width=0.79\textwidth]{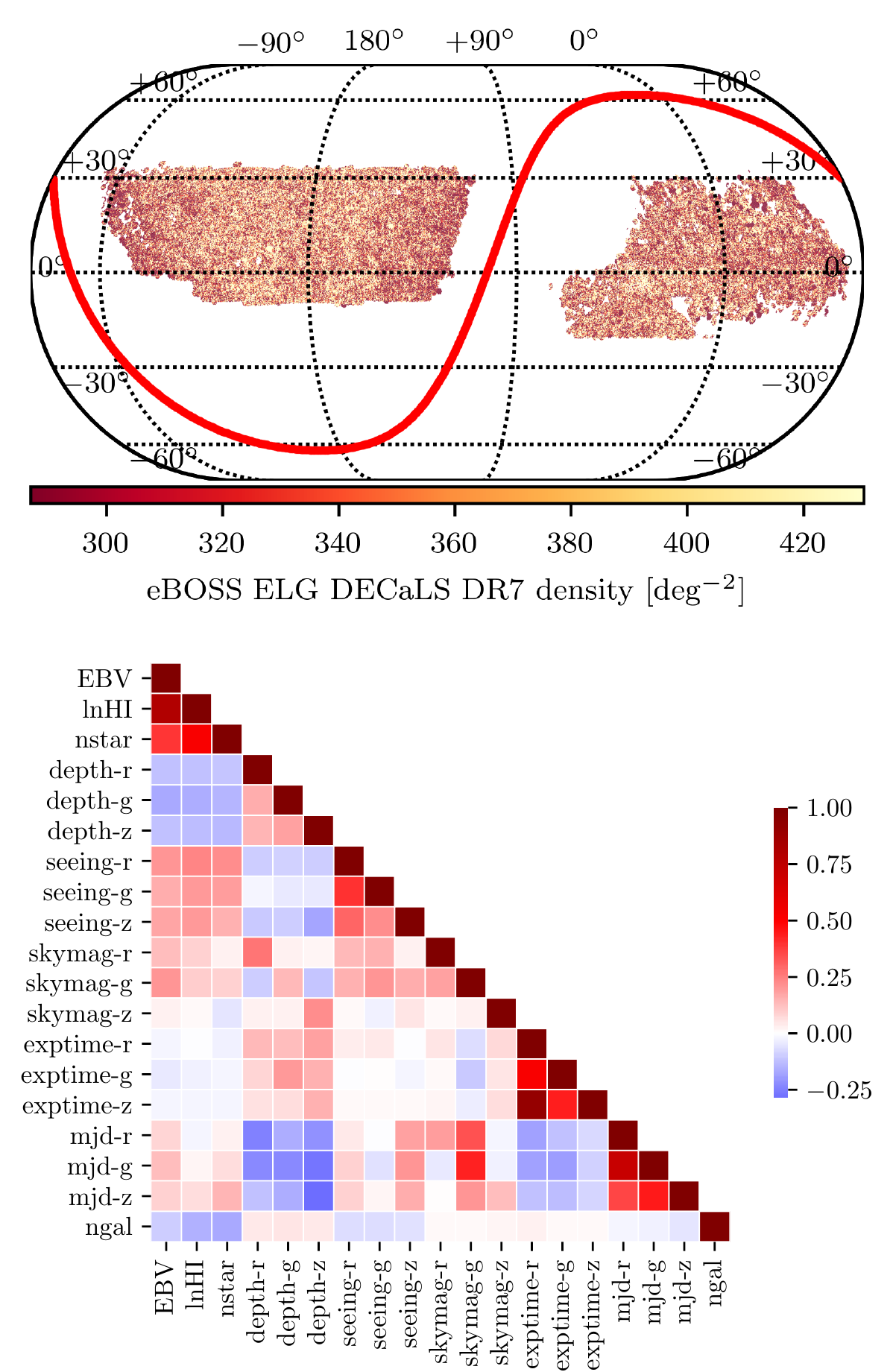}
    \caption{\textit{Top panel}: the pixelated density map of the eBOSS-like ELGs from DR7 after correcting for the completeness of each pixel (see eq., \ref{eq:comp}) and masking based on the survey depth and completeness cuts, see eq.,  \ref{eq:depth_cuts}. The solid red curve represents the Galactic plane. This figure is generated by the code described in \url{https://nbviewer.jupyter.org/github/desihub/desiutil/blob/master/doc/nb/SkyMapExamples.ipynb}. \textit{Bottom panel}: the color-coded Pearson correlation matrix between each pair of the DR7 imaging attributes.}
    \label{fig:eboss_dr7}
\end{figure*}

As an exploratory analysis, we use the Pearson correlation coefficient (PCC) to assess the linear correlation between the data attributes. For two variables $X$ and $Y$, PCC is defined as,
\begin{equation}\label{eq:pcc}
\rho_{X, Y} = \frac{cov(X, Y)}{\sqrt{cov(X,X)cov(Y,Y)}},
\end{equation}
where $cov(X,Y)$ is the covariance between $X$ and $Y$ across all pixels. In Fig.~\ref{fig:eboss_dr7}, we show the observed galaxy density after the pixel completeness (i.e., fracgood $f_{\rm pix}$) correction in the top panel and the correlation (PCC) matrix between the DR7 attributes as well as the galaxy density (\textit{ngal}, the bottom row) in the bottom panel. These statistics indicate that Galactic foregrounds, such as stellar density $nstar$, neural hydrogen column density \lnHI, and Galactic extinction $EBV$, are moderately anti-correlated with the observed galaxy density. Each of these maps traces the structure of the Milky Way and the anti-correlation with $ngal$ implies that, for example, closer to the Galactic plane where the extinction and stellar density are high, there is a systematic decline in the density of galaxies we selected in our sample. The top-left corner of Fig.~\ref{fig:eboss_dr7} shows that these three imaging attributes are strongly correlated with each other. Likewise, the negative correlation of $ngal$ with $seeing$ indicates that as $seeing$ increases the detection of ELGs becomes more challenging. On the other hand, we find a positive correlation between $ngal$ and $depth$s, which can be explained by the fact that as the depth decreases, e.g., we cannot observe fainter objects, the number of galaxies decreases as well.\\

This matrix overall demonstrates that the correlation among the imaging variables is not negligible. For instance, in addition to the aforementioned correlation among the Galactic attributes, there is an anti-correlation between the MJD and depth values. Likewise, there is an anti-correlation between the seeing and depth values. The complex correlation between the imaging attributes causes degeneracies, and therefore, complicates the modeling of systematic effects, which cannot be ignored and needs careful treatment.

\begin{table}
    \centering
    \caption{The statistics of the DR7 imaging attributes used in this paper. Due to the non-Gaussian nature of the attributes, we report the mean, 15.9-, and 84.1-th percentile points of the imaging attributes.}
    \label{tab:meanstats}
    \begin{tabular}{lccr} 
        \hline
        \hline
        \textbf{Imaging map} & 15.9\% &  mean & 84.1\% \\
\hline    
EBV [mag]                      &      0.023 &      0.048 &      0.075 \\ 
ln(HI/cm$^{2}$)                &      46.67 &      47.21 &      47.71 \\ 
\hline    
depth-r [mag]                  &      23.46 &      23.96 &      24.33 \\ 
depth-g [mag]                  &      23.90 &      24.34 &      24.55 \\ 
depth-z [mag]                  &      22.57 &      22.93 &      23.23 \\ 
\hline    
seeing-r [arcsec]              &       1.19 &       1.41 &       1.61 \\ 
seeing-g [argcsec]             &       1.32 &       1.56 &       1.78 \\ 
seeing-z [arcsec]              &       1.12 &       1.31 &       1.51 \\ 
\hline    
skymag-r [mag/arcsec$^{2}$]    &      23.57 &      23.96 &      24.39 \\ 
skymag-g [mag/arcsec$^{2}$]    &      25.06 &      25.39 &      25.80 \\ 
skymag-z [mag/arcsec$^{2}$]    &      21.72 &      22.04 &      22.38 \\ 
\hline    
exptime-r [sec]                &      138.8 &      480.7 &      551.2 \\ 
exptime-g [sec]                &      213.3 &      680.6 &      642.2 \\ 
exptime-z [sec]                &      261.4 &      651.6 &      658.1 \\ 
\hline    
mjd-r [day]                    &    56599.3 &    57232.7 &    57953.3 \\ 
mjd-g [day]                    &    56856.3 &    57358.1 &    57956.3 \\ 
mjd-z [day]                    &    56402.4 &    57005.0 &    57447.3 \\
    \end{tabular}
\end{table}

%% file: sections/methodology.tex
\section{Methodology}\label{sec:method}
\subsection{Observed galaxy density}
In our methodology, we treat the mitigation of imaging systematics as a regression problem, in which we aim to model the relationship between the observed galaxy density (\textit{label}) and the imaging attributes (\textit{features}) that are the potential sources of the systematic error. Note that we do not include the positional information as \textit{input} features since we do not want the mitigation to fit the cosmological clustering pattern. The solution of the regression then would provide the predicted mean galaxy density (i.e., in the absence of clustering or shot-noise) solely based on the imaging attributes of that location. We use this predicted galaxy density as the \textit{survey} selection mask to be applied to any observed galaxy map in the attempt to eliminate the systematic effects and therefore isolate the cosmological fluctuation. Below we describe our procedure.\\

In this paper, we focus on the multiplicative systematic effects. The observed number of galaxies within pixel $i$ can be expressed in terms of the true number of galaxies $n_{i}$ and the contamination model $\mathcal{F}(\textbf{s}_{i})$ as 
\begin{equation}\label{eq:ngal_fs}
    n_{i}^{o}(\textbf{s}_{i}) = n_{i} \mathcal{F}(\textbf{s}_{i}), 
\end{equation}
where \textbf{s}$_{i}$ is a vector representing the imaging attributes \textbf{s} of pixel $i$, and the contamination model $\mathcal{F}(\textbf{s}_i)$ is an unknown function representing the systematic effects which could be either a linear, non-linear, or a more complex combination of the imaging attributes. Multiplicative systematics are associated with obscuration and area-loss due to foreground stellar density, Galactic extinction, etc. On the other hand, additive systematics are associated mostly with stellar contamination, as described in \citet{myers2007clustering, ross2011ameliorating, 2012ApJ...761...14H, prakash2016sdss, 2016MNRAS.455.4301C}. When averaged over many pixels, the effect of additive systematics can be absorbed into the constant term of the multiplicative model $\mathcal{F}$, assuming there is no correlation between the imaging maps and the true galaxy density field. The modeling of $\mathcal{F}(\textbf{s}_{i})$ can be approached by a wide variety of techniques, ranging from the traditional methods based on multivariate functions to non-parametric and non-linear models based on machine learning or deep learning such as random tree forests and neural networks \citep{breiman2001random, geurts2006extremely}.\\

The cosmological information is contained in the true overdensity that is given by
\begin{equation}
\delta_{i} = n_i/(f_{{\rm pix},i}\bar{n})-1,\label{eq:overden}
\end{equation}
accounting for the pixel completeness where $\bar{n}$ is the `true' average number of galaxies. Then,
\begin{equation}
n_{i}^{o}= f_{{\rm pix},i}\overline{n}~(1 + \delta_{i})~ \mathcal{F}(\textbf{s}_{i}).\label{eq:nobsi}
\end{equation}
This $n_{i}^{o}/f_{{\rm pix},i}$ is equivalent to the observed $ngal$ aforementioned.
Since we do not know the true average number density $\overline{n}$ of the data, we estimate $\bar{n}$ from the average of the observed galaxy field,
\begin{equation}\label{eq:nbar}
\hat{\overline{n}} = \frac{\displaystyle\sum_{i} n^{o}_{i}  }{\displaystyle\sum_{i} f_{{\rm pix},i}},
\end{equation}
and treat $\hat{\overline{n}} \equiv \overline{n}$.
Due to the finite volume of our sample, $\hat{\overline{n}} \neq \overline{n}$ even in the absence of systematic effects. This imposes the well-known integral constraint effect on any clustering analysis. We further ignore any systematic effect on $\hat{\overline{n}}$ due to the fact we use $n^{o}_{i}$; that is, Eq.~\ref{eq:nbar} converges to $\overline{n}$ only when $\sum_{i} f_{{\rm pix},i}= \sum_{i} f_{{\rm pix},i}\mathcal{F}_i$. 
In this sense we are modeling the relative systematic effect without necessarily determining the accurate `true' $\overline{n}$. We will use simulated results to test our methodology, and the analysis applied to the simulations with a limited footprint will be subject to the similar finite-volume and systematic effects on $\hat{\overline{n}}$, thus providing a fair comparison and means to catch any obvious problem with this approximation.\\

Finally, we define the normalized galaxy density per pixel $t_{i}$,
\begin{equation}\label{eq:nnbar}
    t_{i}(\textbf{s}_{i}) \equiv \frac{n_{i}^{o}(\textbf{s}_{i})}{f_{{\rm pix},i}\hat{\overline{n}}} = (1+\delta_{i}) ~ \mathcal{F}(\textbf{s}_{i}).
\end{equation}
With this definition, we can estimate the unknown contamination model $\hat{\mathcal{F}}$ (or selection mask) by modeling the dependence of $t_{i}$ on $\textbf{s}_{i}$. When averaged over many spatial positions, the cosmological fluctuations $\delta_i$ will be averaged out and therefore the observed $t_i$ averaged across many pixels with the same imaging attribute, should only be a function of \textbf{s} and return $\mathcal{F}$: 
\begin{equation}\label{eq:tfs}
    <t_{i}(\textbf{s}_{i})>_{i} \simeq <{\mathcal{F}}(\textbf{s}_{i})>_{i} = {\mathcal{F}}(\textbf{s}).
\end{equation}
The inverse of the selection mask which is equivalent to the photometric weights ($wt^{{\rm sys}}$) in other studies can therefore be used to remove the systematic effects from the observed galaxy number map (cf. Eq. \ref{eq:ngal_fs}),

\begin{align}\label{eq:wsys}
\hat{n}_i = \frac{n^o_i}{\hat{\mathcal{F}}}=n^o_i wt^{{\rm sys}}_{i}.
\end{align}
In the following we describe how we obtain $\hat{\mathcal{F}}$ using different regression approaches, e.g., neural networks and multivariate linear functions. From now on, the terms \textit{features} and \textit{label} associated with each data point refer to \textbf{s} and $t$ of each HEALPix pixel, respectively. 

\subsection{Mitigation with Neural Networks}\label{subsec:MethodNN}
\begin{figure}
\centering
\includegraphics[width=0.4\textwidth]{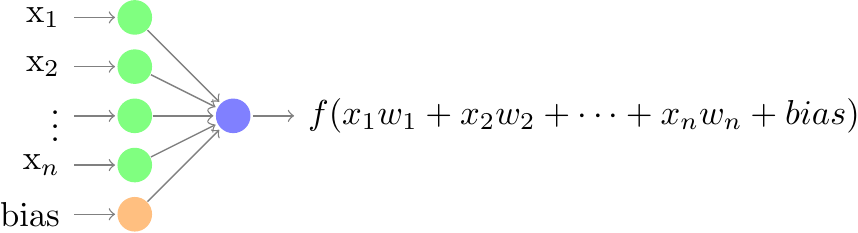}
\caption{A schematic diagram of a single neuron with the activation function $f$. The neuron takes a set of inputs \textbf{x}=$(x_{1}, x_{2},...,x_{n})$, multiplies each of them by its associated weight \textbf{w}=$(w_{1}, w_{2},...,w_{n})$, and sums the weighted values and a thresh-hold (or a constant offset) which is called \textit{bias}, to form a pre-activation value, $z=\sum_{i=1}^{n}x_{i}w_{i} + bias$, which is a linear process. The neuron then transforms the pre-activation $z$ to the output using the activation function $f(z)$, which is where the nonlinear process can enter. }\label{fig:perceptron}
\end{figure}

We will apply {\it fully connected feed forward neural networks} in order to tackle our regression problem. Fig. \ref{fig:perceptron} illustrates a schematic diagram of a neuron, the building block of a neural network, which generates the output based on a linear combination of the inputs followed by a nonlinear transformation, the activation function $f$.\\

\begin{figure}
\centering
\includegraphics[width=0.45\textwidth]{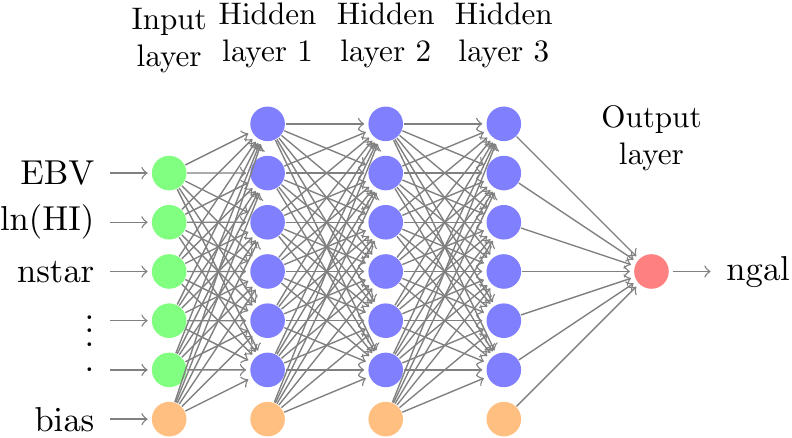}
\caption{A schematic illustration of a fully connected feed forward neural network with the imaging attributes in the input layer, three hidden layers of six neurons in the middle, and a single neuron on the output layer, as an example. The blue-colored neurons have non-linear activation functions, while the red-colored neuron lacks any activation function. In reality, we employ the validation procedure to choose the best number of hidden layers while keeping the total number of hidden layer neurons fixed at 40 (i.e., approximately twice the number of imaging attributes in this study).}\label{fig:ffnn}
\end{figure}

Fig. \ref{fig:ffnn} illustrates the architecture of a fully connected feed forward neural network with the imaging attributes in the input layer, three hidden layers of six non-linear neurons in the middle, and a single neuron without any activation function in the output layer, as an example. The \textit{bias} neuron in each layer is shown in orange and is analogous to the intercept in linear regression. The output of the neural network will be an estimation of the contamination model $\hat{\mathcal{F}}$ (see Eq. \ref{eq:nnbar}). If we use the identity function as the activation function (e.g., $f(z)=z$), regardless of the number of hidden layers, the neural network is equivalent to a linear model. This means that our methodology is a generalization or an extension of the conventional linear mitigation methods. The modeling capabilities of neural networks depend on the number of hidden layers, type of non-linear activation function and the number of neurons in each hidden layer \citep[see e.g.][]{cybenko1989approximation,hornik1989multilayer,funahashi1989approximate, tamura1997capabilities, huang2003learning, lin2017does, rolnick2017power}.\\

We use the rectifier $f(z) = \text{max}(0, z)$ as the activation function for the hidden layer neurons, which alleviates the `vanishing gradient' problem
~\citep[see e.g.,][]{nair2010rectified,glorot2011deep,krizhevsky2012imagenet, dahl2013improving,montufar2014number}. \\

\begin{figure}
\centering
 \includegraphics[width=0.35\textwidth]{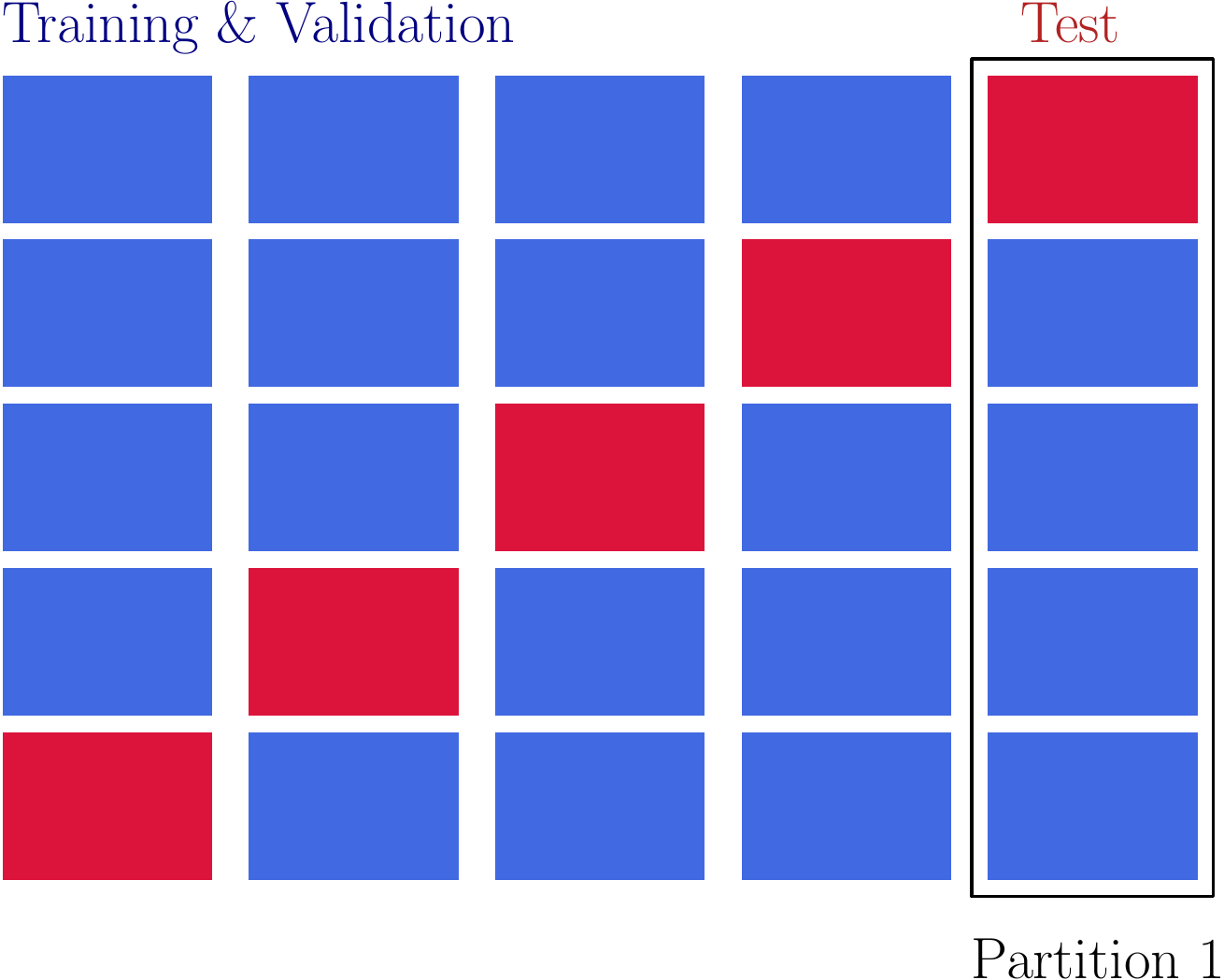}
 \caption{A visualization of the five-fold permutations of data-split. The data is randomly split into five equal-size folds, and by permutation of the folds we construct five partitions of data. For each partition/permutation, three folds are assigned to the training set, one fold for the validation set, and the remaining one fold for the test set: therefore, 60\% training, 20\% validation, and 20\% test sets. Each column represents a partition. The test folds are shown in red, while the training and validation folds are shown in blue. The key points are : 1) The folds are not contiguous (in RA, DEC) as may be implied by this cartoon. 2) There is no overlap between the training, validation, and test folds within a partition. 3) One can reconstruct the entire data by merging the test folds from the five partitions.}
\label{fig:5fold}
\end{figure}

We utilize  $k$-fold cross-validation with $k=5$ folds/sub-groups to train the parameters, tune the hyper-parameters, and to estimate the predictive performance of the neural network. As illustrated in Fig. \ref{fig:5fold}, we randomly split the entire pixel data (187,257 pixels) into five folds and construct the training, validation and test data sets out of these five folds; three folds are assigned to the training set, one fold is assigned to the validation set, and the remaining one fold is assigned to the test set. A specific assignment of the five folds to these three sets forms one `partition'. We construct five different partitions such that each fold is used once as test fold. This $k$-fold cross-validation scheme ensures that a test example is never used for training or tuning.\\

We standardize (i.e., renormalize) the label and features of the training, validation, and test folds using the mean and standard deviation of the label and features of the corresponding training fold (i.e., similar to Tab.~\ref{tab:meanstats}, but for the training set of each partition). We initialize the biases to zero and sample the weights of each layer randomly from a normal distribution whose variance is inversely proportional to the number of neurons of the previous layer \citep[see e.g.,][]{he2015delving}. Using the training fold, we utilize the adaptive gradient descent with momentum \citep[\textit{Adam},][]{kingma2014adam} to update the parameters of the neural network with batches of size $N_{{\rm batch}}$. Thus, the entire training set is split into $N_{{\rm batch}}$ batches and a gradient update is applied for each batch. One training epoch corresponds to processing the $N_{{\rm batch}}$ batches once \citep[for more details on the training procedure, we refer the reader to see e.g.,][]{ruder2016overview}. The hyper-parameters of \textit{Adam}, specifically the moments decay rates and the tolerance, are fixed as follows: $\beta_{1}$ = 0.9, $\beta_{2}$ = 0.999 and $\epsilon$ = 10$^{-8}$. The default learning rate of 0.001 will be tuned using the validation data.\\ 

The network is trained to minimize the following cost function:
\begin{equation}\label{eq:cost}
    \text{J} = \frac{1}{N_{{\rm batch}}}\sum_{i}^{N_{{\rm batch}}}f_{\rm pix, i}~[t_{i} - \hat{\mathcal{F}_{i}}]^{2} + \frac{\lambda}{2} ~ ||w||^{2} ,
\end{equation}   
where the first term is the Mean Squared Error (MSE) weighted with $f_{\rm pix,i}$, and the second term is the L2 regularization term, used to penalize higher weight magnitudes and a larger number of neurons~\citep{hoerl1970ridge}. The strength of the L2 penalty term is controlled via the regularization scale $\lambda$. The network is trained for a number of training epochs, $N_{{\rm epoch}}$, although to avoid unnecessary training, we implement the early stopping technique with the tolerance of 1.e-4 and patience of 10, i.e., the training terminates if the validation MSE does not improve more than the tolerance within the last 10 epochs.\\

\subsubsection{Backward feature elimination}
    
The input features are highly correlated as shown in Fig.~\ref{fig:eboss_dr7}, and therefore the 18 maps probably contain redundant information.
We apply \textit{backward feature elimination} (feature selection) to remove the redundant input features in order to reduce the noise in the prediction as well as to protect the cosmological information by avoiding too much freedom in modeling. We find that reducing the dimension of the input features, i.e., the imaging attributes, is an essential step to avoid over-fitting and regressing out the cosmological clustering.\\

\begin{figure}
    \centering
    \includegraphics[width=0.4\textwidth]{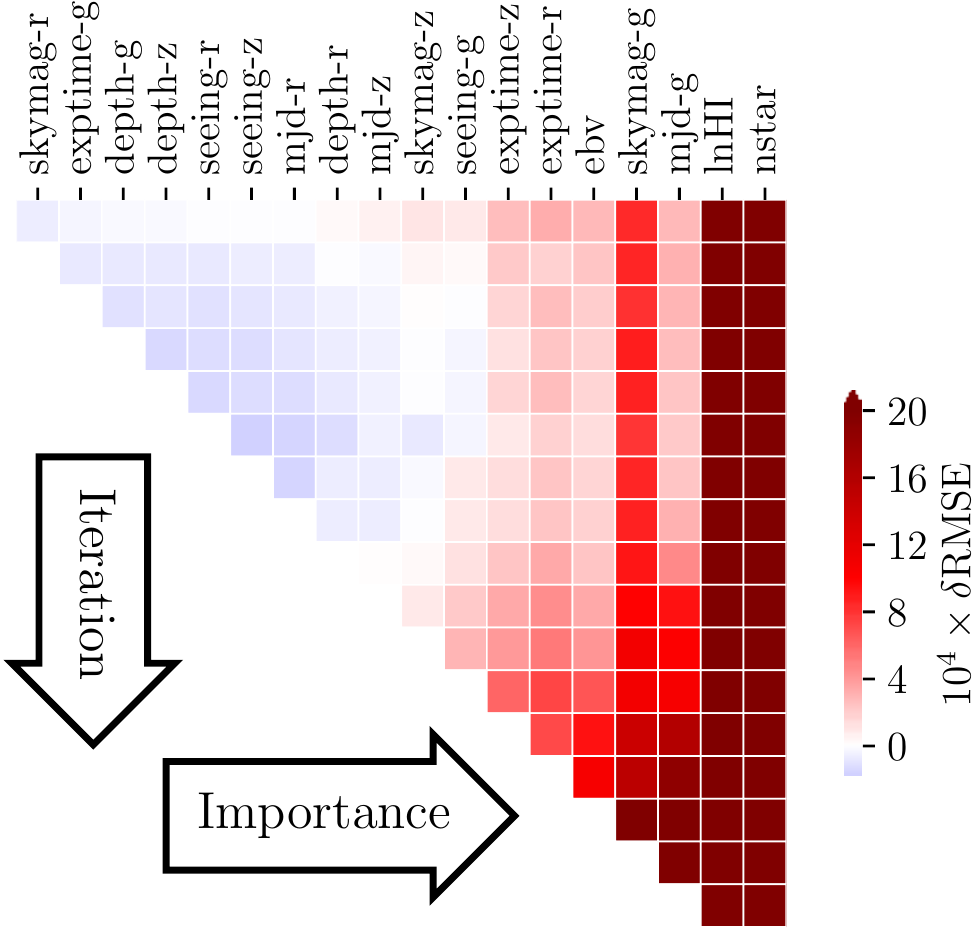}
    \caption{Feature importance of DR7 based on backward feature elimination for the first partition, as an example. This process iteratively removes the feature whose removal produces the largest decrease in the validation RMSE (i.e., the greatest improvement in fitting) until no decrease is observed. After the first iteration of removal (the top row), the removal of $skymag-r$ decreased the validation RMSE the most and therefore $skymag-r$ is removed. In the second iteration (the second row), removing $exptime-g$ decreased the validation RMSE the most and therefore it is removed. However, in the ninth iteration, the removal of $mjd-z$ did not decrease the validation RMSE and therefore the feature selection stops here, passing the rightmost ten features to the neural network regression. As a result of the process, the importance increases from left to right, and the rightmost ten maps in the figure ($ebv$, $nstar$, $logHI$, $seeing-g$, $skymag-g$, $skymag-z$, $exptime-r$, $exptime-z$, $mjd-g$, $mjd-z$) are the ones that worsen the validation RMSE when being removed from the input layer.}
    \label{fig:dr7ablation}
\end{figure}

\begin{figure}
    \centering
    \includegraphics[width=0.4\textwidth]{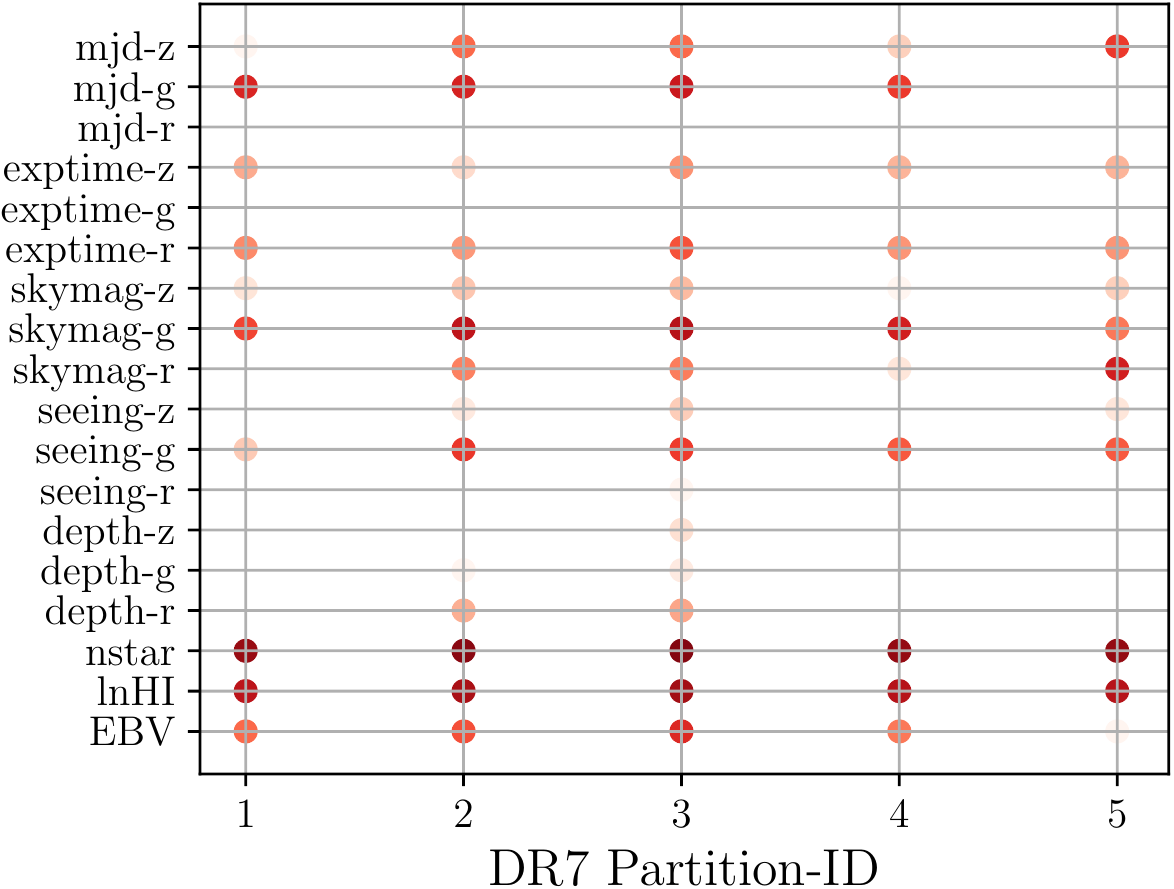}
    \caption{Important imaging maps identified by the backward feature elimination (\textit{feature selection}) procedure for the five partitions used for DR7. A darker color of a point within each partition represents a more important attribute identified by the feature selection procedure. Note that $nstar$ is identified as the most important attribute in all partitions, i.e., across the footprint.}
    \label{fig:dr7ablation2}
\end{figure}

We perform the feature selection for each partition separately. Initially, we train a linear model on all 18 input features with the following hyper-parameters: the initial learning rate of 0.001, batch-size of 1024, L2 regularization scale of zero, for 500 training epochs with early stopping. We record the validation MSE as a baseline criterion. Then we eliminate one input feature and train the linear model on the remaining 17 features. This trained linear model is applied to the validation set. The input feature whose removal has produced the highest decrease in validation MSE (i.e., the highest improvement in fitting) is permanently eliminated, leaving only 17 features for training. Note that if the feature contained useful information on the systematic effects, removing the feature would have made the fit worse. We repeat the regression using 17 input features, and so on, removing one feature at each iteration until either the validation MSE does not decrease relative to the baseline or all input features are removed. Fig. \ref{fig:dr7ablation} shows the result of the backward feature elimination procedure for the first partition of DR7, ranking the input features based on their importance from left to right. This result supports the trends seen in the exploratory analysis in \S~\ref{sec:data} which indicated strong linear correlations with the stellar density, Galactic extinction, and hydrogen column density in the data (see the correlation matrix in the bottom panel of Fig. \ref{fig:eboss_dr7}). The color gradient indicates the relative change in validation Root Mean Squared Error (RMSE) when that particular feature is removed with reference to the baseline. We note that the order of removal is not the same as, for example, the color gradient order of the attributes in the first iteration. We believe it is because, as we remove the redundant features, the relevant importance of the remaining input features changes due to the complex correlations between the removed features and the remaining ones. The remaining features for each of the five partitions are shown in Fig. \ref{fig:dr7ablation2}. The attributes $lnHI$ and $nstar$ are commonly identified as the most important features and then $ebv$, $seeing-g$, $skymag-g$, $skymag-z$, $exptime-r$, $exptime-z$, $mjd-z$ are commonly identified for all 5 partitions.\\

\subsubsection{Hyper-parameter tuning, training, and testing}\label{subsubsec:hyperparam}

We train the hyper-parameters for each partition separately. At each training epoch and for each choice of hyper-parameters, we apply the trained network on the validation fold. We adjust the hyper-parameters accordingly such that the validation MSE is minimized. Our neural network has the following five hyper-parameters: number of hidden layers; number of training epochs $N_{{\rm epoch}}$; L2 regularization scale $\lambda$; batch size $N_{{\rm batch}}$; Adam's learning rate. We tune one hyper-parameter at a time. To find the optimal learning rate, we monitor the behavior of the cost function during training. We observe that a learning rate of 0.001 leads to a smoothly decreasing cost function vs. training epochs. We train the network for up to $N_{\rm epoch} = 500$ epochs although we implement the early stopping technique with the inertia (or patience) of 10 and the tolerance of 1.e-4: this means the training will be stopped if none of the last 10 epochs achieved a smaller relative error reduction with respect to the minimum validation error, within the tolerance. For the number of hidden layers, we try the following architectures, in which the total number of hidden neurons is fixed at 40 (i.e. roughly twice the number of the features) except for the linear model:\\~\\
$[0]$  : no hidden layers \\
$[40]$ : one hidden layer of 40 neurons \\
$[20, 20]$ : two hidden layers of 20 neurons on each \\
$[20, 10, 10]$ : three hidden layers of 20, 10 and 10 neurons \\
$[10, 10, 10, 10]$ : four hidden layers of 10 neurons\\~\\
After finding the best number of layers, we proceed to tune $\lambda$ by trying powers of 10, e.g., 0.001, 0.01, ..., 1, ..., 1000. Finally, we adjust $N_{\rm batch}$ by trying powers of 2, e.g., 128, 256, ... , 4096. The optimal set of the hyper-parameters for each partition is summarized in Tab. \ref{tab:hparams}.\\

\begin{table}
    \centering
    \caption{The best hyper-parameters for each partition of DR7.}
    \label{tab:hparams}
    \begin{tabular}{lccr} 
        \hline
         & number of layers & $\lambda$ & $N_{\rm batch}$ \\
        \hline
        Partition 1 & [20, 20] & 0.001 & 4096\\
        Partition 2 & [20, 10, 10] & 0.001 & 512\\
        Partition 3 & [20, 10, 10] & 0.001 & 1024 \\
        Partition 4 & [20, 10, 10] & 0.001 & 512 \\
        Partition 5 &  [20, 10, 10] & 0.001 & 2048\\
    \end{tabular}
\end{table}

\begin{figure*}
         \centering
         \includegraphics[ width=0.8\textwidth]{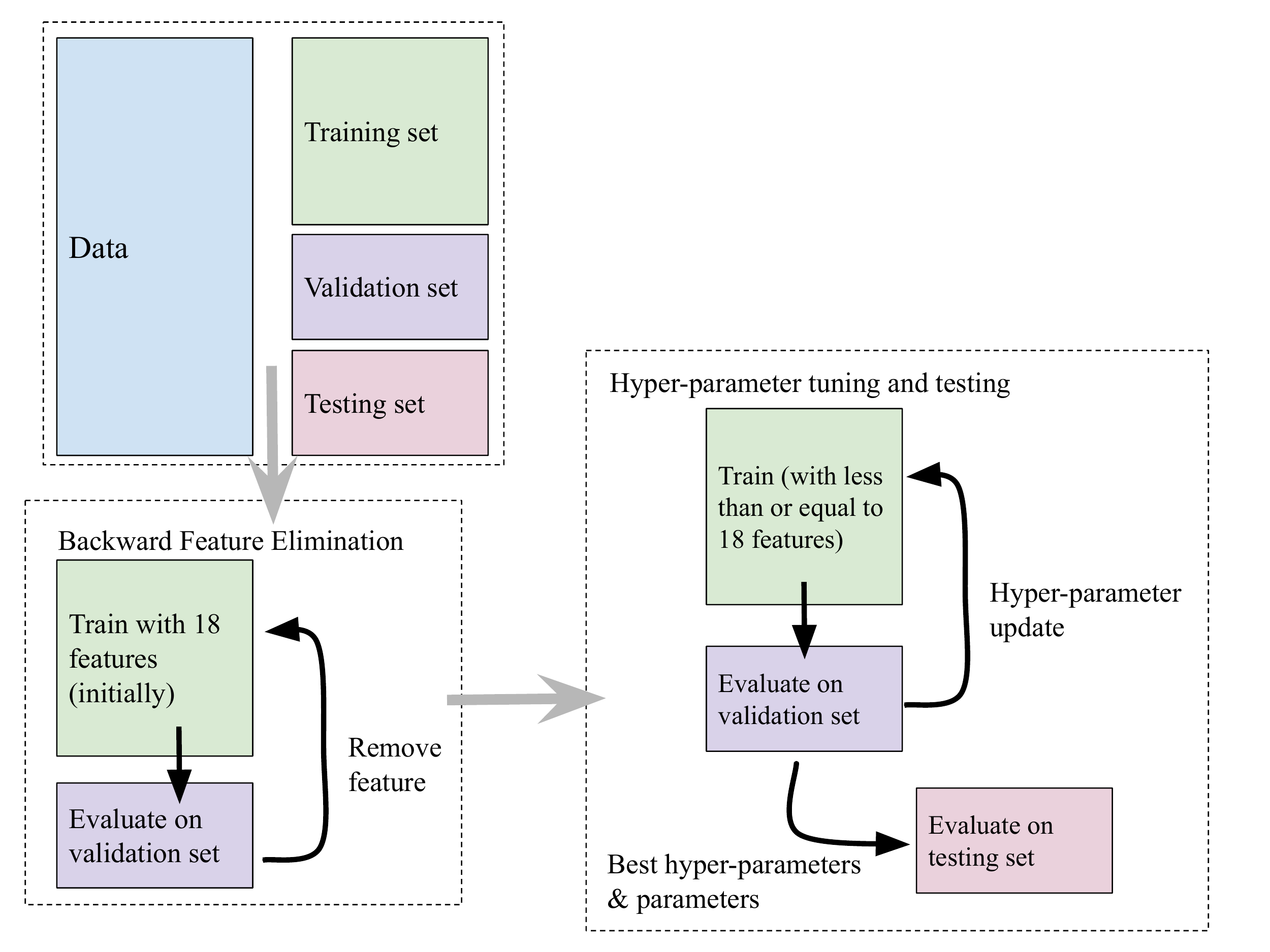}
         \caption{A flow-chart of the backward feature elimination and hyper-parameter tuning for each partition. This entire process is performed five times for each of the five partitions/permutations such that the entire footprint is covered by aggregating the different testing folds. }
         \label{fig:pipeline}
\end{figure*}

Once the grid search procedure identifies the best performing hyper-parameters out of the predefined ranges introduced in Section \ref{subsubsec:hyperparam}, the network is trained with these hyper-parameters for 10 independent runs, each one with a different initialization of the weights and biases, and then applied on the test set. We compute the median of the predicted test label from the 10 runs and aggregate the results over the 5 different partitions to construct the map of the predicted label ($\hat{\mathcal{F}}$) for the entire footprint. For our default method, the backward feature elimination is conducted for each partition and reduces the number of input features before the hyper-parameter training step, as illustrated in Fig.~\ref{fig:pipeline}.  This process is performed for each partition separately, each partition using a different fold as the test set, until the entire footprint is covered through the 5 test folds. The flow of the feature selection, hyper-parameter tuning, and testing is summarized in Fig. \ref{fig:pipeline}.\\

\subsection{Mitigation with Multivariate Linear Functions}
We use linear and quadratic polynomial functions to model the normalized galaxy density dependence on the imaging attributes (Eq.~\ref{eq:tfs}), as the benchmark approaches to be compared with the neural network. Unlike the proposed neural network method, no regularization or dimensionality reduction is performed, and all data are used to train the parameters of the regression models. Despite a lack of any deliberate machinery against over-fitting, we note that over-fitting is less likely to be an issue for this method since the size of the data is much greater than the number of the fitting parameters. Nevertheless, we tried splitting the sample into 60\% of the training data to derive the best fit linear coefficients and 20\% of the test set (i.e., the same training and test sample size to have a fair comparison with the neural network) to apply the derived coefficients and permuted five times until the test set covers the entire footprint. We find that such a split does not change the results of the linear regression. On the other hand, the limited flexibility of its parameterized form could be a weakness of this method and we believe it is responsible for the differences that the neural network method makes in the comparison presented in Section \ref{sec:mitigateDR7}.\\

The contamination model from Eqs.~\ref{eq:nnbar} and ~\ref{eq:tfs} can be estimated via a multivariate linear function in terms of the standardized imaging attributes (\textbf{s}) as,

\begin{equation}\label{eq:multvar}
\hat{\mathcal{F}}(\textbf{s}|b_{0},\alpha) = b_{0} + \sum_{m=1}^{M}\sum_{k=1}^{18} \alpha_{mk} (\frac{s_{k}-\overline{s_{k}}}{\sigma_{k}})^{m} ,
\end{equation}
where $M$ is the maximum power law index equal to 1 and 2 for the linear and quadratic polynomial model, respectively; the constants $\overline{s_{k}}$ and $\sigma_{k}$ are respectively the mean and standard deviation of the k'th imaging quantity, $s_{k}$ (cf. Tab. \ref{tab:meanstats}). The parameters $b_{0}$ and $\alpha_{mk}$ are the intercept and the corresponding coefficients for each term, respectively, which are tuned by minimizing the weighted sum of the squared errors. The output of the regression is applied as the selection mask on the observed galaxy density to eliminate the systematic effects (see Eq., \ref{eq:wsys}).\\

\subsection{Angular Clustering Statistics}\label{subsec:ang_clustering}
\subsubsection{One point statistics}
In the absence of the systematic effects, the galaxy density field should be statistically independent from the imaging attributes and will only depend on the cosmological fluctuations while an individual dataset/mock will be subject to chance correlations within the statistical error. When averaged over many spatial positions, the cosmological fluctuations will be averaged out and therefore the average density should be equal to the mean density once the survey footprint is accounted for. A deviation from the mean density as a function of the imaging attributes, therefore, is indicative of the average dependence of the observed galaxy density on the corresponding imaging attributes. To assess the level of the contamination in the data, we compute the histogram of the spatially averaged galaxy density vs the imaging attributes. For each of the system attribute, $s_k$, we prepare 20 bins. For each bin of $s_{k}$, we have:

\begin{equation}\label{eq:nnbar_stat}
    \frac{\overline{n}(s_{k})}{\overline{n}_{tot}} \equiv \frac{1}{\hat{\overline{n}}}~ \frac{\displaystyle\sum_{s_{k}\leq s_{k,i}<s_{k}+\Delta s_{k}} n^{o}_{i}}{\displaystyle\sum_{s_{k}\leq s_{k,i}<s_{k}+\Delta s_{k}} f_{\rm pix, i}},
\end{equation}
where the indices $i$ and $k$ respectively represent the pixel index, and the systematic index; $\Delta s_{k}$ is the bin width arranged for different $s_k$ bins such that each bin contains almost the same amount of effective area, in an attempt to suppress fluctuations in $\overline{n}(s_{k})/\overline{n}_{tot}$ due to small number statistics. We estimate the error bars using the Jackknife resampling of 20 non-contiguous subsamples of pixels within each imaging attribute bin ~\citep[see e.g.][]{ross2011ameliorating}:

\begin{equation}\label{eq:error_jack}
    \sigma^{2}_{{\rm Jack}}(s_{k}) = \frac{19}{20}\sum_{j=1}^{20} \left[    \frac{\overline{n}(s_{k})}{\overline{n}_{tot}} -     \frac{\overline{n}_{j}(s_{k})}{\overline{n}_{tot}}\right]^{2}  ,
\end{equation}
 where $\overline{n}_{j}(s_{k})/\overline{n}_{tot}$ is computed over the entire sample when the $j$'th Jackknife region is excluded.
As a result of the adjusted $\Delta s_{k}$, the level of $\sigma^{2}_{{\rm Jack}}(s_k)$ is almost the same for all $s_k$. After mitigation of the systematic effects, one expects that the corrected density field is independent of the imaging attributes, i.e., $\overline{n}(s_{k})/\overline{n}_{tot}$ being consistent with unity.

\subsubsection{Two-point clustering statistics}
The two-point clustering statistic measures the spatial correlation of the galaxy density and has been the main statistic for extracting the cosmological information from galaxy surveys. We use the angular auto and cross two-point clustering statistics of the galaxy density field as well as of the imaging attributes to estimate the impact of the potential systematics on the cosmological clustering signal and to examine the effectiveness of the mitigation techniques tested in this paper. For pixel $i$, we calculate the galaxy overdensity $\delta_{i}$ using Eq.~\ref{eq:overden} and the fluctuation of a given imaging attribute $\delta^{s}_{i}$ as
\begin{align}\label{eq:delta}
    \hat{\delta}^{s}_{i} &= \frac{s_{i}}{\hat{\overline{s}}} - 1 ,
\end{align}
where $\hat{\overline{s}}$ is the mean of each imaging attribute weighted with $f_{{\rm pix},i}$,
\begin{align}\label{eq:deltasysbar}
\hat{\overline{s}} = \frac{\sum_i f_{{\rm pix},i} s_i}{\sum_i f_{{\rm pix},i}}
\end{align}
following \citet{ross2011ameliorating}.\\

By definition, Eqs.~\ref{eq:nbar} and \ref{eq:delta} ensure that the following integral of the observed quantity over the entire footprint vanishes:
\begin{align}\label{eq:ic}
    \sum_{i} \hat{\delta}_{i}~f_{{\rm pix},i} = 0,
\end{align}
for both the galaxy as well as imaging attribute fluctuations. We utilize both the angular correlation function and angular power spectrum to extract the cosmological information from the galaxy density field. While our mitigation efficiency is evaluated based on the angular power spectrum, we also inspect the angular correlation function to make sure that the systematics are mitigated in that estimator as well, since both estimators are commonly used in the clustering analysis and they are complementary to each other given the limited range of data.

\begin{figure}
        \centering
        \includegraphics[width=0.4\textwidth]{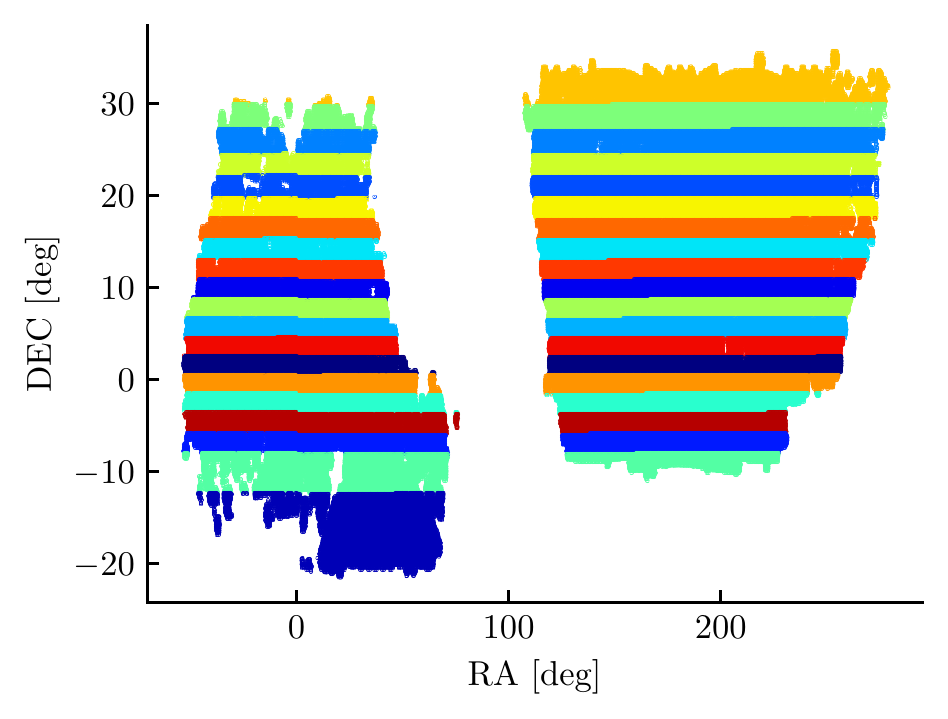}
        \caption{Twenty equal-area contiguous regions used to estimate the Jackknife errorbars for the 2D clustering statistics.}
        \label{fig:jackknifes}
\end{figure}

\begin{itemize}
    \item Angular Correlation Function: we employ the HEALPix-based estimator to compute the angular correlation function which, for a separation angle $\theta$, is defined as \citep[see e.g.][]{scranton2002analysis,ross2011ameliorating}
    \begin{align}
        \omega^{p,q} (\theta) = \frac{\displaystyle\sum_{ij} \hat{\delta}^{p}_{i} \hat{\delta}^{q}_{j} \Theta_{ij}(\theta) f_{\rm pix, i}f_{\rm pix, j}}{\displaystyle\sum_{ij} \Theta_{ij}(\theta) f_{\rm pix, i}f_{\rm pix, j}},\label{eq:ancor}
    \end{align}
    where $p=q$ gives an auto correlation function estimator, $p\neq q$ gives a cross correlation function estimator, and $\Theta_{ij}$ is one when two pixels $i$ and $j$ are separated from each other within $\theta$ and $\theta+\Delta\theta$, or zero otherwise. Note that our estimator weighs each pixel overdensity with $f_{{\rm pix}, i}$ since the pixels with a greater complete area coverage should have a higher signal to noise. Such weight is straightforwardly corrected by the denominator in Eq.~\ref{eq:ancor} unlike its conjugate estimator (i.e., the power spectrum). Since our overdensity map resolution is limited by the pixel size, we set the $\Delta\theta$ to be the resolution of a pixel ($\sim$ 0.23 deg). \\

    \item Angular Power Spectrum : one can conveniently expand a coordinate on the surface of a sphere in terms of spherical harmonics or, if azimuthally symmetric, Legendre polynomials. We define the following estimator for expanding the galaxy overdensity:
    \begin{equation}
        \hat{\delta}_{i} = \sum_{\ell=0}^{\ell_{{\rm max}}}\sum_{m=-\ell}^{\ell} a_{\ell m} Y_{\ell m}(\theta_{i}, \phi_{i}),
    \end{equation}
    where $\theta, \phi$ represent the polar and azimuthal angular coordinates of pixel \textit{i}, respectively. The cutoff at $\ell=\ell_{{\rm max}}$ assumes that the signal power is not significant for modes $\ell>\ell_{{\rm max}}$. We define the following spherical harmonic (SH) transform estimator of overdensity ($\hat{\delta}$) over the total number of non-empty pixels $N_{{\rm pix}}$:
   
    \begin{equation}
        \hat{a}_{\ell m} = \frac{4\pi}{N_{{\rm pix}}} \sum_{i=1}^{N_{{\rm pix}}}  \hat{\delta}_{i}~f_{\rm pix, i}~ Y^{*}_{\ell m}(\theta_{i}, \phi_{i}),
    \end{equation}
    where $^{*}$ represents the complex conjugate, and we again down-weight the overdensity in pixel \textit{i} by the completeness ($f_{\rm pix, i}$). Due to the survey window function implicit in the sum over the non-empty pixels and explicit in $f_{\rm pix, i}$, our estimator would not return unbiased estimates of the SH coefficients, unless the window function effect is corrected for, and also the expected orthogonalities between different SH modes would not hold. Nevertheless, we define the angular power spectrum estimator as the average of the magnitude of SH coefficients over $m$:
    \begin{equation}\label{eq:pusedocell}
        \hat{C}^{p,q}_{\ell} = \frac{1}{2\ell +1} \sum_{m=-\ell}^{\ell} \hat{a}^{p}_{\ell m} \hat{a}^{q*}_{\ell m},
    \end{equation}
    where $p=q$ gives an auto power spectrum, $p\neq q$ gives a cross power spectrum between the galaxy density and the imaging attributes. In order to compute the angular power spectrum, $C_{\ell}$, we make use of the ANAFAST function from HEALPix \citep{gorski2005healpix} with the third order iteration of the quadrature to increase the accuracy\footnote{We refer the reader to \url{https://healpix.sourceforge.io/pdf/subroutines.pdf}, page 104.}. 
    Unlike in the angular correlation function, we do not attempt to correct for the survey window function/survey mask effect in the angular power spectrum both for DR7 and the mocks. We rather calculate the window effect on the theoretical models of power spectrum in Appendix \ref{app:windowfunction}. For the mock test, we use the angular power spectrum observed in the mocks without the contamination model, i.e., the `Null' case, as our baseline to compare with different mitigation methods.\\ 
\end{itemize}
We use the Jackknife resampling technique with 20 equal-area contiguous regions, as shown in Fig. \ref{fig:jackknifes}, to estimate the error-bars on $\omega (\theta)$ and $C_{\ell}$ (see Eq. \ref{eq:error_jack}).\footnote{We use the mocks without imaging systematics (null mocks in Section \ref{subsec:surveymocks}) to compare the errors from the Jackknife subsamples of one mock with the errors among 100 full  DECaLS-like mock footprints, and we find that the former is greater than the latter on ell $<$ $\sim$ 10 by a factor of 2-3, possibly due to the dispersion in the survey window function of the Jackknife samples. For a real survey the observational condition also varies across the footprint. Therefore, with Jackknife errors, the significance of any improvement on systematics treatment will be conservatively assessed.} For both the mock and real datasets, we also utilize the cross power spectra between the galaxy density and various imaging maps to evaluate the performance of the mitigation. In order to estimate the significance of the contamination in $\hat{C}^{g,g}_{\ell}$ (or $\omega^{g,g} (\theta)$) before and after mitigation, we calculate $[\hat{C}^{g,s_k}_{\ell}]^2/\hat{C}^{s_k,s_k}_{\ell}$ (or $[\omega^{g,s_k}(\theta)]^2/\omega^{s_k,s_k}(\theta)$) as a proxy.~\footnote{These quantities would be the true level of contamination to $\hat{C}^{g,g}$ if the contamination model is linear and systematics are independent of one another~\citep{ashley2012MNRAS,2012ApJ...761...14H}.}

\subsection{Survey Mocks}\label{subsec:surveymocks}
Imaging systematics tend to affect the clustering signal mainly on large scales \citep{myers2007clustering, huterer2013calibration} and the distribution of galaxies on large scales at moderately low redshift can be well-approximated by a log-normal distribution~\citep{1991MNRAS.248....1C}. We therefore believe that log-normal mocks would be sufficient for the purpose of validating our systematic mitigation techniques. We use the \textit{Nbodykit} package \citep{hand2017nbodykit} to generate one hundred log-normal cubic mocks with the box-side length of $5274 \ihMpc$ and $1024^3$ mesh cells, with the input power spectrum matched to the linear power spectrum  at $z=0.85$ based on the \textit{Planck 2015} cosmology~\citep{ade2016planck} (i.e., flat $\Lambda$CDM with $\Omega_m=0.3089 \pm 0.0062$, $H_{0}=67.74 \pm 0.46$, $\sigma_8 = 0.8159 \pm 0.0086$), with the galaxy bias of 1.5 and the volume density of $1.947\mathrm{e}{-4}\trihMpc$ \citep[see e.g.][]{Raichoor2017MNRAS.471.3955R}. Then, we use the \textit{make\_survey} package \citep{white2013mock} to sub-sample the mock galaxies based on the NGC eBOSS ELG redshift distribution in \citet{Raichoor2017MNRAS.471.3955R} with the redshift cut of  0.55 $<$ z $<$ 1.5 and to transform the cubic mocks into survey-like mocks. We do not include redshift-space distortions (RSD) in the mocks as we believe that the systematics mitigation efficiency does not depend on  the presence of RSD.\\

\begin{figure}
    \centering
    \includegraphics[width=0.4\textwidth]{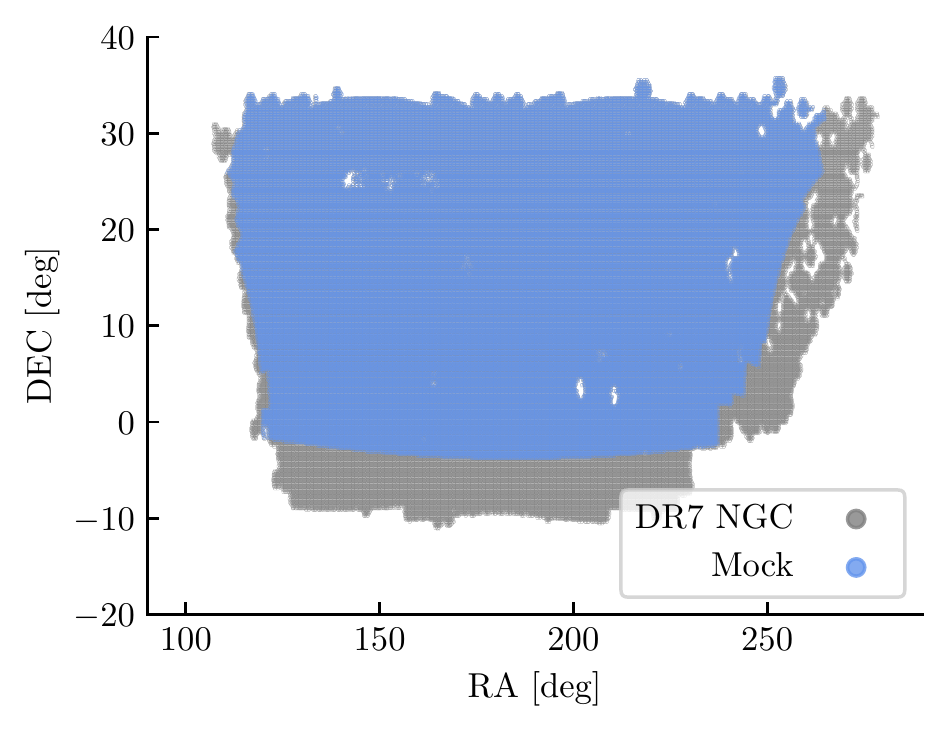}
    \caption{The projection of the mock footprint (blue) onto the North Galactic Cap of the DR7 footprint (gray). With this projection, the imaging attributes from the real data are assigned to the mocks.}
    \label{fig:mock_on_dr}
\end{figure}

The survey mocks are then projected onto the two-dimensional sky using HEALPix and overlaid on the NGC footprint of DR7 to be assigned with the DR7 imaging attributes. Fig. \ref{fig:mock_on_dr} illustrates the resulting projection of a simulated survey mock and DR7. Note that the mock footprint (89,672 pixels) is smaller than DR7 (187,257 pixels) almost by a factor of 2. We only use the pixels of the mock that have the DR7 imaging attributes available. The holes (e.g., RA and DEC around 200 and 5 deg) are the pixels that do not have the imaging attributes from the real data. In order to account for the mock survey footprint, we distribute 2,500 random points per deg$^{2}$ within the mock footprint and derive the completeness map for the mocks.

\subsubsection{Null Mocks}
Our goal is to develop a systematics treatment methodology that maximally removes the systematic effects while minimally removes the true cosmological signal. The two aspects may not be simultaneously accomplished, in a way that depends on the signal, noise, and the correlation between the imaging maps and the true galaxy density. In our paper we choose to prioritize losing minimal cosmological information over maximally removing systematics. One way of ensuring this is to check if the mitigation method returns the true clustering in the presence of contaminations, which will be tested using the contaminated mocks. Another way is to check if the mitigation method correctly makes a null operation on the clustering in the absence of contaminations, returning $\hat{\mathcal{F}}\simeq 1$. To this end, we utilize the 2D projected mocks without introducing any modulation due to imaging attributes in the galaxy density fields. Henceforth, we call this set of simulations, \textit{null} mocks.\\

Fig.~\ref{fig:ngal_hist} shows the pixel distribution histograms of the number of galaxies per pixel of the mocks in comparison to that of DR7 on the common footprint. The average $ngal=7.0/{\rm pixel}$ of the null mocks is smaller than $ngal=13.3/{\rm pixel}$ of the DR7 data (even after accounting for the 5\% loss due to tiling completeness and 27\% loss due to the redshift range, as stated in \citet{Raichoor2017MNRAS.471.3955R}). We believe that the difference in $ngal$ is due to the different \textit{clean photometry} criteria applied to the ELG selection in \citet{Raichoor2017MNRAS.471.3955R} and to the targets in this paper. The standard deviation of $ngal$ of the null mocks is 3.0, which is smaller than 4.6 of the DR7 ELGs.

\subsubsection{Contaminated Mocks}
We modulate the mock galaxy density fields using imaging attributes of DR7 and generate the contaminated mocks with additional random noise. The modulation is done based on the best fit coefficients of the imaging attributes and their covariances for $\mathcal{F}$ (Eq.~\ref{eq:nnbar}) that we derived from DR7 using our fiducial linear regression model.\\

\begin{figure}
    \centering
    \includegraphics[width=0.4\textwidth]{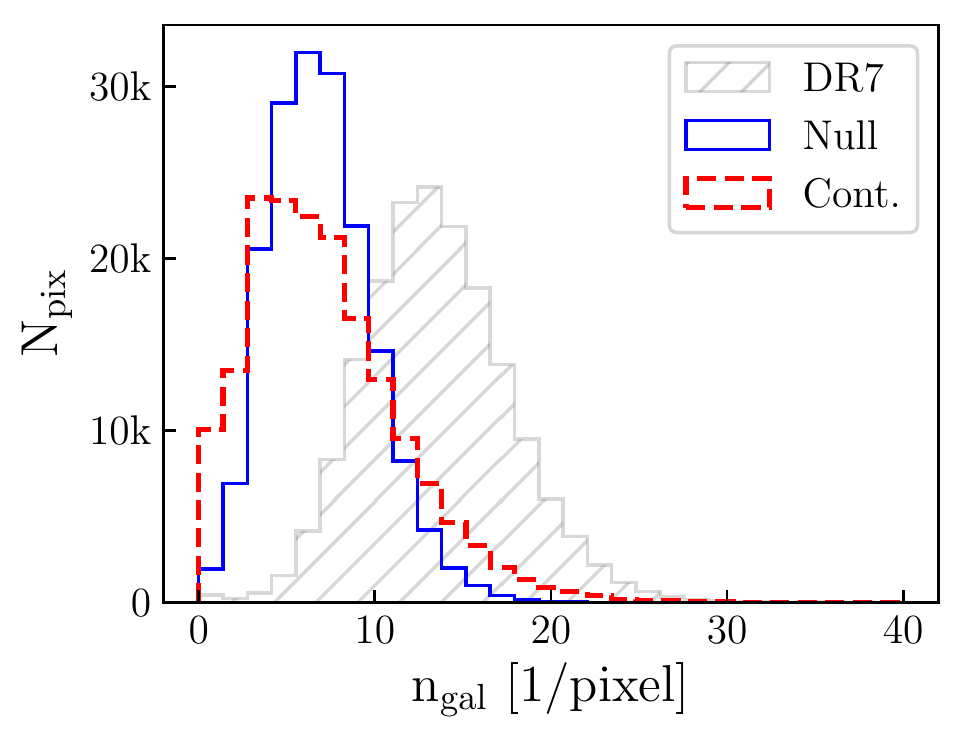}
    \caption{Histogram of the number of galaxies per pixel for DR7 (hatched grey), null (solid blue), and contaminated (dashed red) mocks. All distributions are corrected for the pixel completeness $f_{\rm pix}$. The DR7 distribution is scaled down to account for 5\% tiling completeness, and 27\% to 0.55 < reliable redshift <1.5. The residual difference between the mocks and  DR7 shown here might be due to differences between the \textit{clean photometry} criteria applied to eBOSS target selection and those we apply to DR7.}
    \label{fig:ngal_hist}
\end{figure}

In detail, we pick the 10 imaging attributes, i.e., $EBV$, $nstar$, $lnHI$, $seeing-g$, $skymag-g$, $skymag-z$, $exptime-r$, $exptime-z$, $mjd-g$, $mjd-z$ of DR7, which were selected from the feature selection procedure on one of the partitions, and modulate the mock density field $n$ with $\mathcal{F}$ that is derived from random deviates of the imaging attribute coefficients while accounting for their covariances. Since the measured covariance of such quantities includes both the cosmological fluctuation and the fluctuations due to the imaging attributes, we rescale the measured covariance matrix of the systematics such that the random fluctuation in $ngal$ per pixel due to contamination is at a similar level to the cosmological fluctuation from the null mocks. As a result of the random fluctuation in the contamination model we introduced, some of the pixels will be assigned a negative galaxy number. We drop these pixels from our sample. This removes 3.1\% of the mock footprint, reducing our mock footprint size from 89,672 to 86,875 pixels. We then introduce the Poisson process, i.e., another random variation step, to ensure the modulated galaxy number per pixel is an integer. These two random variation processes increase the noise in the mock datasets such that the standard deviation of $ngal$ of the contaminated mocks ($=4.4$) is almost the same as that of DR7, despite the different average $ngal$ (see Fig.~\ref{fig:ngal_hist}).\\

Therefore our mock contamination is simpler than DR7 in that we adopted a linear model, which is chosen purposely since we do not want to give a priori advantage to our neural network method and also since all methods are capable of reproducing the linear model. Meanwhile, this setup is more challenging than the DR7 data since the mitigation is conducted in the presence of a greater level of noise. Note that, while we included only 10 dominant imaging attributes in the contamination, the remaining attributes in the DR7 data are correlated with these 10 attributes and therefore with the modulated galaxy density. All of the mitigation methods in the following mock test will be challenged to deal with such indirect correlations among the 18 attributes. Note that the effect of the footprint, i.e., the survey window effect, is the same for both the null mocks and contaminated mocks since we chose to apply the selection function on the galaxies while leaving the randoms intact. Therefore the null mocks serve as the baseline for estimating the level of systematics in the contaminated mocks.

%% file: sections/results.tex
%
%
\section{Results}\label{sec:results}
In this section, we present the measurements of the clustering statistics before and after correcting for the systematic effects for the real dataset as well as the simulated ones. We demonstrate that the neural network is capable of learning more structure in the observed galaxy density field due to its greater flexibility beyond a fixed functional form, and therefore it can eliminate more excess clustering which is believed to be due to the imaging systematics. We then show the performance of the neural network and multivariate linear models when applied to the mock datasets.

\subsection{Mitigating systematics from DR7}
\label{sec:mitigateDR7}
\begin{figure*}
    \centering
    \includegraphics[width=\textwidth]{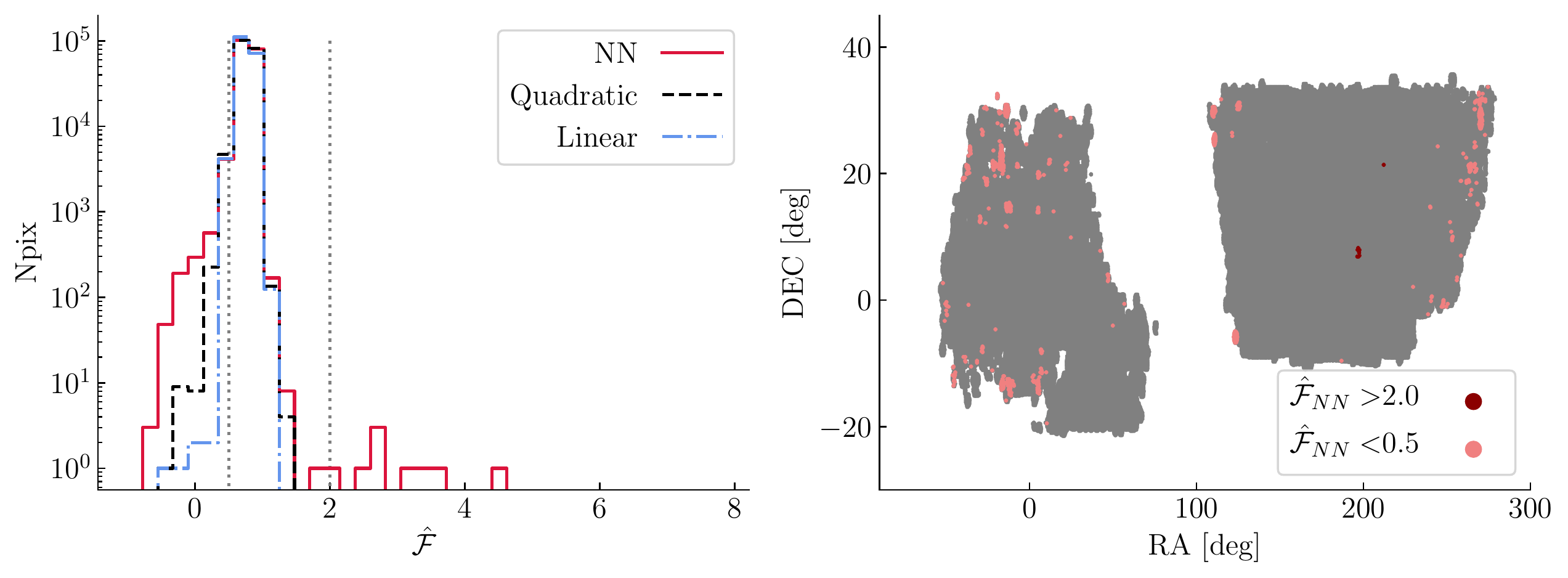}
    \caption{The pixel distribution  of  the  selection  masks for DR7. \textit{Left}: Distribution of the selection masks (i.e., estimates of the contamination model) derived from different regression models. \textit{Right}: Spatial scatter of the pixels we remove from our data due to the extreme values of the neural network selection mask.}
    \label{fig:weights}
\end{figure*}

In the left panel of Fig. \ref{fig:weights}, we show the pixel distribution histograms of the selection masks from the three different regression models we consider in this paper. While all three models show fairly consistent selection masks for most of the pixels (note the logarithmic scaling of $Npix$), the neural network method (solid red curve) returns extended tails due to a higher representation flexibility associated with its nonlinear nature. We remove pixels with $\hat{\mathcal{F}} < 0.5$ or $> 2.0$ from our data to avoid too aggressive selection correction since we believe none of these methods can be accurate enough for such a long baseline extrapolation. These pixels account for 1.0\% of the original data (from 187,257 to 185,781 pixels). In the right panel, we show the spatial distribution of the removed pixels in the case of the neural network selection mask. \\

\begin{figure*}
    \centering
    \includegraphics[width=\textwidth]{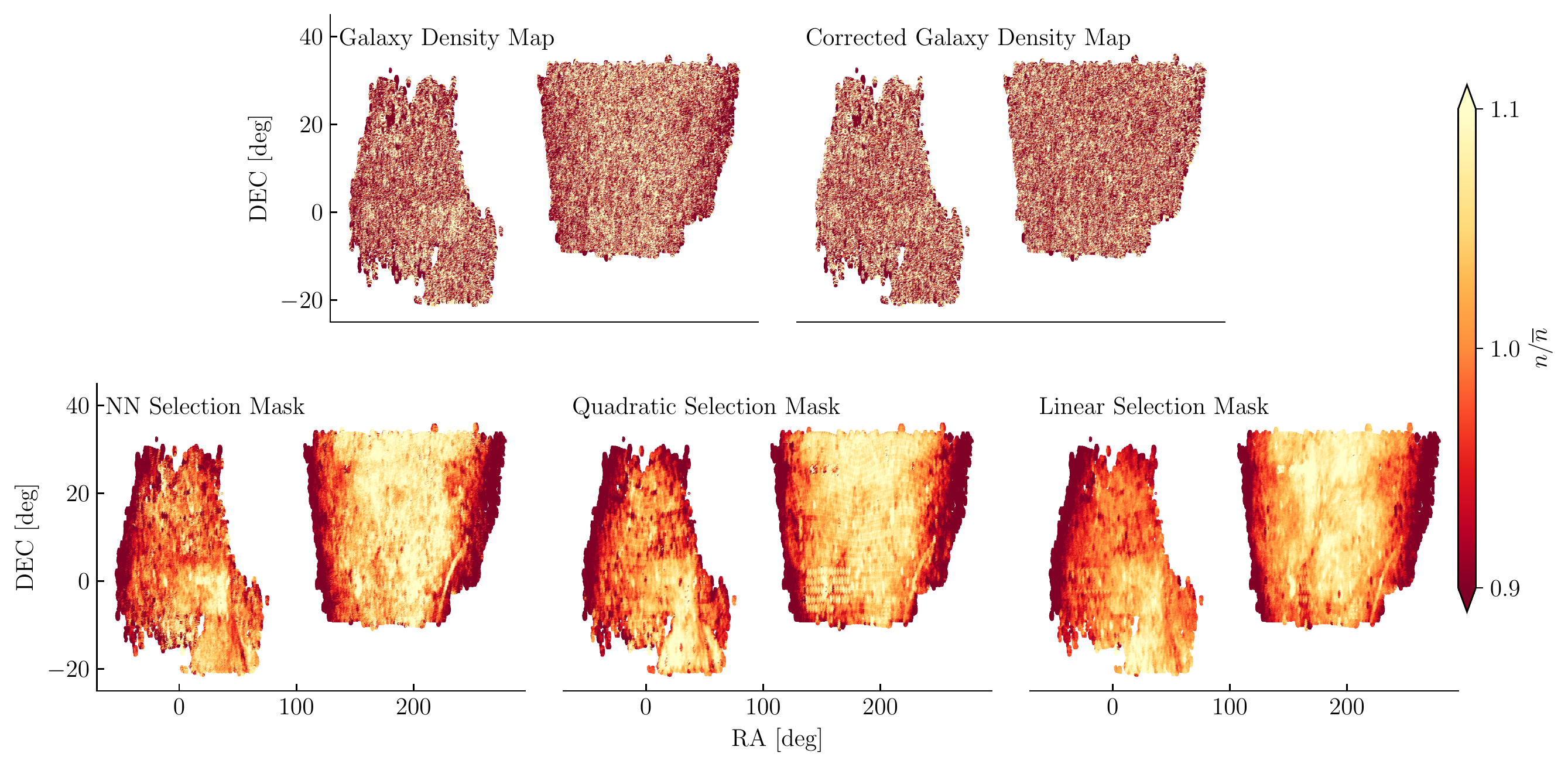}
    \caption{\textit{Top}: The normalized observed galaxy density of  DR7 before and after systematics treatment using the neural network selection mask, respectively, from left to right. \textit{Bottom}: the selection masks from the Neural Network, quadratic, and linear polynomial models, respectively from left to right. All three selection masks are able to capture the behavior that the galaxy density systematically drops at the footprint boundaries i.e., high extinction regions.}
    \label{fig:density_selection}
\end{figure*}

Fig. \ref{fig:density_selection} illustrates the spatial distribution of the observed galaxy density before (top left) and after correction (top right) using the neural network selection mask. The bottom panels show the neural network selection mask used for the correction (left) in comparison to the masks derived from the linear (middle) and the quadratic polynomial (right) models. All three masks capture a very similar large scale pattern such as the decrease in the galaxy density close to the Galactic midplane, which is consistent with the negative correlation coefficients between the galaxy density and the Galactic extinction, hydrogen column density, or stellar density. On smaller angular scales, the three selection masks show different fluctuation details. In the following analyses, we examine which method returns the least contaminated galaxy density distribution. \\

\begin{figure*}
\centering
\includegraphics[width=0.79\textwidth]{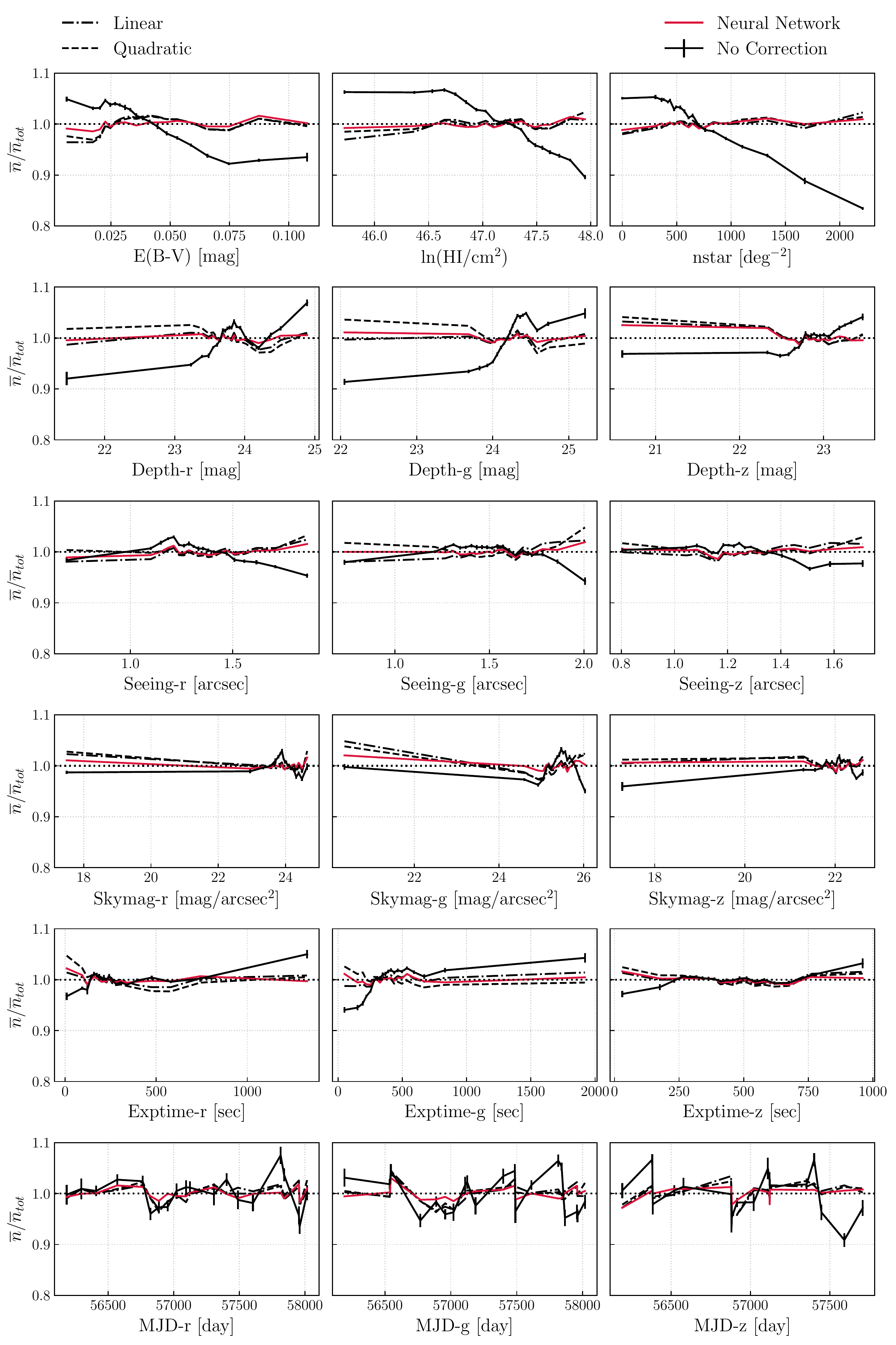}
\caption{The mean number density of the DR7 galaxies as a function of the potential systematics. The solid black curve shows the result before mitigation (\textit{no correction}); the solid red curve is for the result after correcting with the neural network selection mask; the dot-dashed and dashed black curves represent mitigations with the linear and quadratic polynomial selection masks, respectively. The error bars are estimated using the Jackknife resampling of 20 non-contiguous subsamples of pixels within each imaging attribute bin (a total of 20 bins per attribute) and are shown only for one case. This plot again shows that the Galactic foregrounds such as the stellar density introduce a systematic trend in the galaxy density, which indicates a significant contamination by our own galaxy before mitigation. Such systematic trends are mostly removed with any of the three mitigation methods. \label{fig:nnbar}}
\end{figure*}

First, once the systematic effects are corrected for, the mean galaxy density should be independent of the imaging attributes. In Fig. \ref{fig:nnbar} we show the mean number density of galaxies as a function of the imaging attributes. Again, different bins are set to include the same effective pixel area and therefore have the same sampling error. The solid black curve shows the galaxy mean density before correction and the solid red shows the result after correction with the selection mask of the neural network model. The dot-dashed curve represents the correction using the linear polynomial model and the dashed curve is for the quadratic polynomial model. The errorbars are computed using 20 Jackknife sub-samples and shown on only one case for clarity. Similar to what we found from the feature selection procedure, the stellar density, Galactic extinction, and HI density exhibit the strongest dependence before correction. After correction, all three methods return the fractional galaxy density close to unity. To quantify the deviation from unity, we report the $\chi^{2}$ statistics in Tab.~\ref{tab:chi2} while ignoring the covariance between the different bins and different imaging attributes. Overall, the neural network achieves the smallest deviation from unity which indicates its highest efficiency in reducing the systematic effects. Ideally we would like to have residual contamination less than the statistical error. Figure \ref{fig:nnbar} and Table \ref{tab:chi2} implies that we need to further improve the mitigation techniques for future cosmological analyses. In Section \ref{subsec:discussion} we provide a more detailed analysis using the same $\chi^{2}$ statistics and the mocks to quantify the remaining systematics and assess whether or not the data is clean enough.\\

\begin{table}
    \centering
    \caption[Caption]{The $\chi^{2}$ values for the measured mean density of the DR7 galaxies vs imaging systematics, presented in Fig. \ref{fig:nnbar}. This table presents the cumulative values over all bins and all imaging attributes (i.e., $N_{\rm bins}=$20 bins $\times$ 18 attributes) without accounting for the covariance both between the imaging maps and between different bins\protect\footnotemark.}
    \label{tab:chi2}
    \begin{tabular}{lccr} 
        \hline
        Correction scheme & $\chi^{2}$ & $N_{\rm bins}$ & $\chi^{2}/N_{\rm bins}$ \\
        \hline
        None & 20921.633 & 360 & 58.116\\
        Linear & 2588.349 & 360 & 7.190\\
        Quadratic & 2623.006 & 360 & 7.286 \\
        Neural Network & 966.601 & 360 & 2.685\\
    \end{tabular}
\end{table}

\begin{figure*}
\centering
\includegraphics[width=0.78\textwidth]{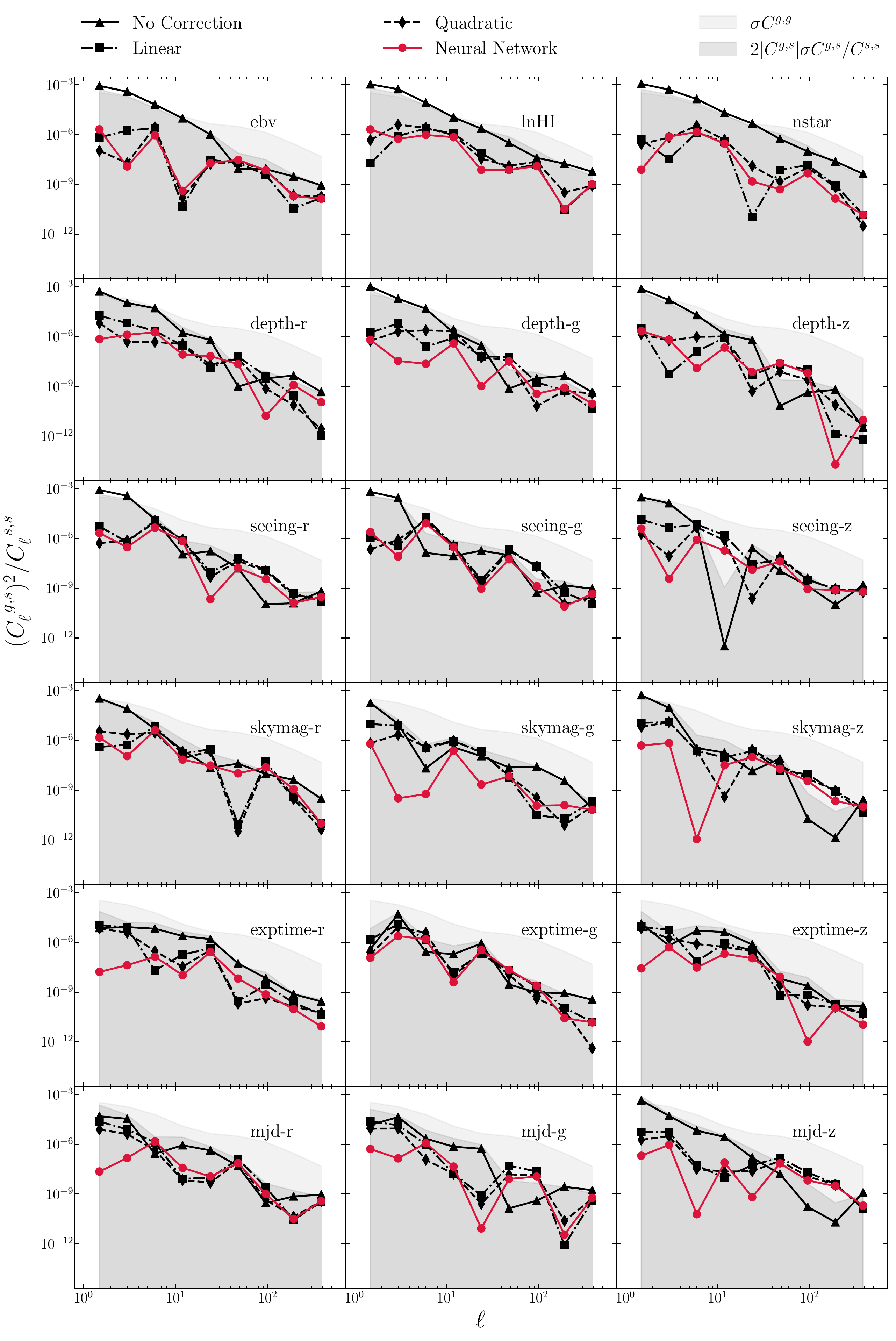}
\caption{The cross power spectrum $\hat{C}^{g,s_k}_{\ell}$ between the DR7 observed galaxy density and the imaging attributes $s_k$ normalized by the auto power spectrum of the imaging attribute $\hat{C}^{s_k,s_k}_{\ell}$. The plotted quantity $[\hat{C}^{g,s_k}_{\ell}]^2/\hat{C}^{s_k,s_k}_{\ell}$ approximately represents the level of contamination to the auto power spectrum of the galaxy density $\hat{C}^{g,g}_{\ell}$. The light and dark grey shaded regions, respectively, show the Jackknife error estimate of $\hat{C}^{g,g}_{\ell}$ and $[\hat{C}^{g,s_k}_{\ell}]^2/\hat{C}^{s_k,s_k}_{\ell}$ with the galaxy density $g$ before mitigation. The black solid curve shows the result before mitigation (\textit{no correction}), while the solid red curve shows the result after correcting for the systematics with the neural network selection mask. The dot-dashed and dashed black curves show the corrected results with the linear and quadratic polynomial model selection masks, respectively. \label{fig:clcross}}
\end{figure*}

\begin{figure*}
\centering
\includegraphics[width=0.79\textwidth]{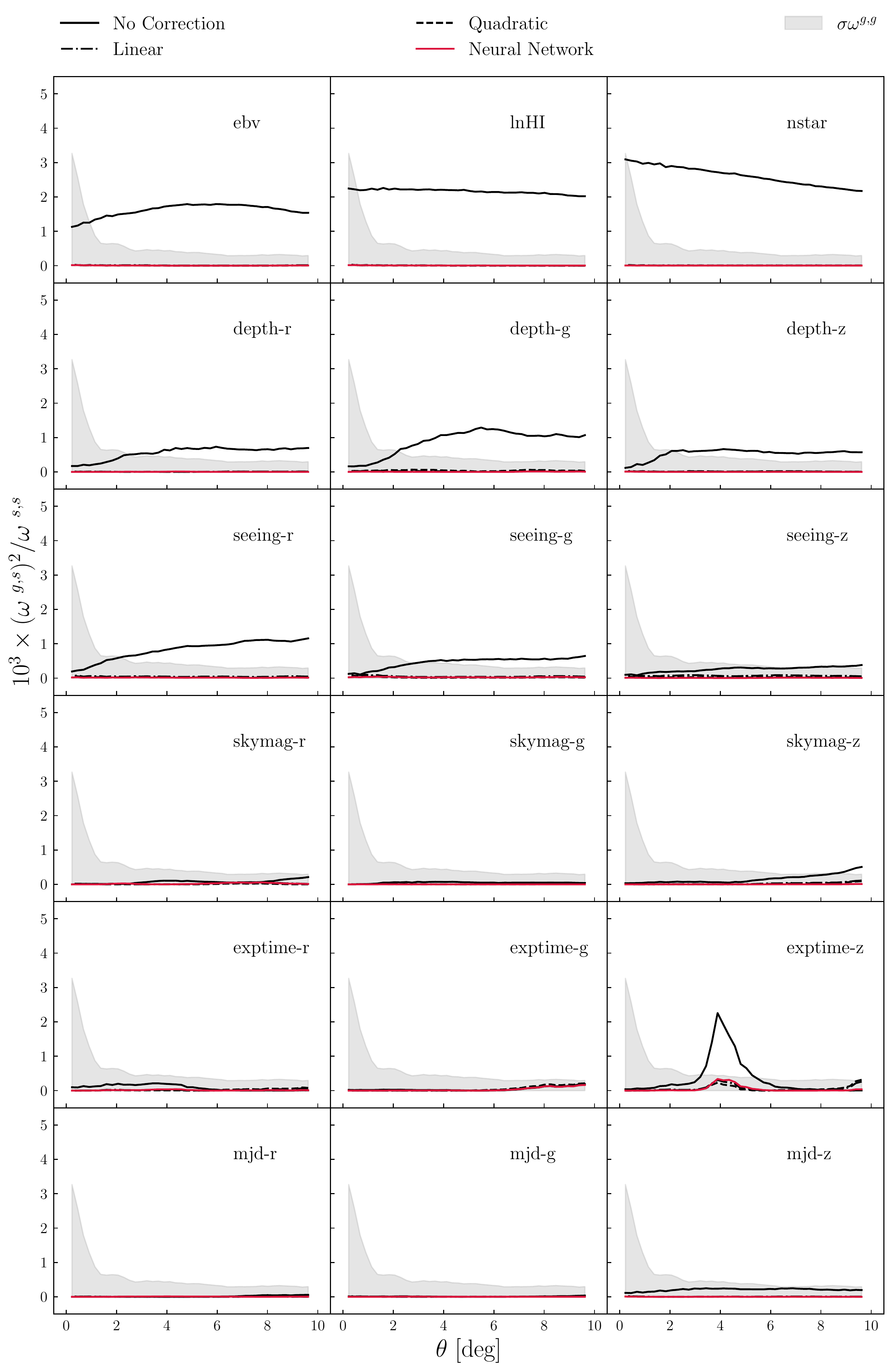}
\caption{The cross correlation function $\omega^{g,s_k}(\theta)$ between the DR7 observed galaxy density and the imaging attributes $s_k$ normalized by the auto correlation function of the imaging attribute $\omega^{s_k,s_k}(\theta)$. The plotted quantity $[\omega^{g,s_k}(\theta)]^2/\omega^{s_k,s_k}(\theta)$ approximately represents the level of contamination to the auto correlation function of the galaxy density $\omega^{g,g}(\theta)$.  The grey shaded region shows the Jackknife error estimate of $\omega^{g,g}(\theta)$  before mitigation. All mitigation techniques are able to reduce the excess clustering signal which is due to the imaging systematics.  \label{fig:xicross}}
\end{figure*}

We next evaluate the performance of different mitigation techniques using the two-point statistics. We first show the cross power spectra between the DR7 observed galaxy density and various imaging attributes in the form of  $[\hat{C}^{g,s_k}_{\ell}]^2/\hat{C}^{s_k,s_k}_{\ell}$ in Fig. \ref{fig:clcross}. Again, this quantity approximately represents the level of contamination from each attribute to the auto power spectrum of galaxy density and we therefore compare this with the uncertainty in the auto as well as cross power spectrum of galaxies (light and dark gray shades) which are estimated using the Jackknife resampling of 20 equal-area contiguous regions (see Fig. \ref{fig:jackknifes}). Similarly, we plot $[\omega^{g,s_k}(\theta)]^2/\omega^{s_k,s_k}(\theta)$ in Fig.~\ref{fig:xicross} to assess the contamination in the auto-correlation function. Fig. \ref{fig:clcross} and \ref{fig:xicross} show significant contamination on large scales from $ebv$, $lnHI$, and $nstar$ compared to the statistical fluctuation estimated from the Jackknife subsampling of the data. The stellar density map shows the highest cross power spectrum with the galaxy density map, which is in agreement with the previous results. Qualitatively, all three mitigation techniques perform well and substantially reduce the cross power below $\ell \sim 30$ and over all separation scales in the cross-correlation function. The neural network method shows a slightly lower cross-power, but this appears to be merely related to the lower amplitude of the corresponding auto galaxy power spectrum compared to the other two cases. We note the spurious peak in the cross-correlation against exptime-z in Fig.~\ref{fig:xicross} near the expected angular location of the BAO feature and such feature necessitates thorough investigations of imaging systematics in analyzing the auto clustering statistics of the spectroscopic data for BAO analysis.\\

\begin{figure*}
\centering
\includegraphics[width=\textwidth]{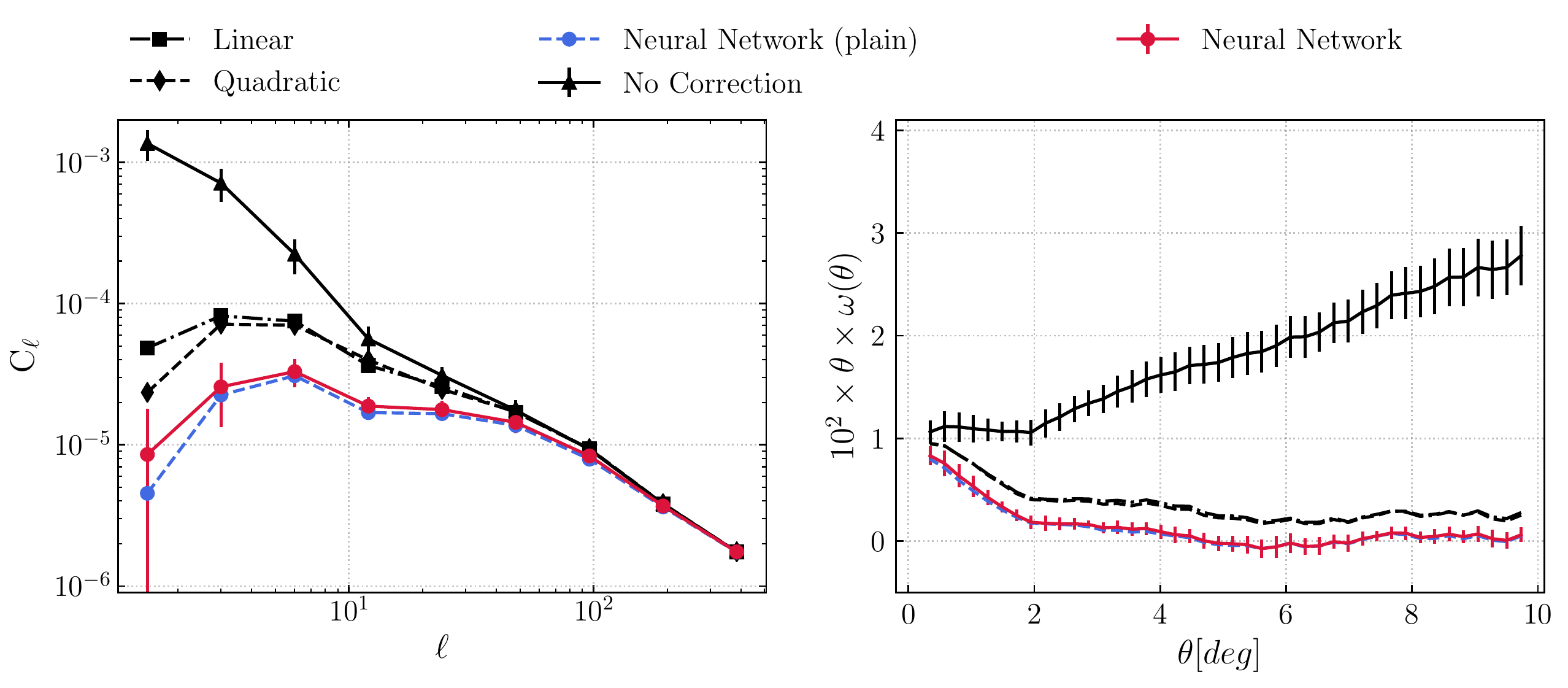}
\caption{Two-point clustering statistics for DR7. \textit{Left}: the measured angular power spectrum without shot-noise subtraction. \textit{Right}: the HEALPix-based angular correlation function. Solid black curves show the measured statistics without correcting for the systematic effects (\textit{no correction}). The dashed and dot-dashed black curves show the statistics after correcting with linear and quadratic polynomial mitigation methods, respectively. The solid red curves show the results after correcting with our default neural network method. The dashed blue curves show the results mitigated with the neural network method but without the feature selection process. The errors are estimated using the Jackknife resampling with 20 contiguous sub-regions and are shown only for a few cases for clarity (see Fig. \ref{fig:jackknifes}).} \label{fig:clxi}
\end{figure*}

We finally present the effect of the imaging attributes before and after mitigation on the auto galaxy clustering statistics. In Fig. \ref{fig:clxi}, we illustrate the measured two-point clustering statistics for DR7; the measured angular power spectrum without shot-noise subtraction is shown in the left panel, and the HEALPix-based angular correlation function is shown in the right panel. The solid black curve shows the measured clustering before mitigation, while the corrected measurements using the traditional linear, quadratic polynomial, and the default neural network models are shown respectively with the black dot-dashed, black dashed, and solid red curves. In the right panel, the linear and the quadratic polynomial mitigation results are indistinguishable and overlaid.\\ 

\footnotetext{Note that the best fit neural network model was applied to the unseen data (i.e., the test set) unlike in the linear and quadratic polynomial models. Nevertheless the neural network method returns the smallest $\chi^{2}$, i.e., the highest efficiency.}

The comparison between the clustering before correction (solid black curves) and  after treatment (solid red, blue, dashed, and dot-dashed curves) suggests that the imaging systematics affect the clustering measurements mostly on large scales, e.g., large separation angles or small multipoles, as expected~\citep[see e.g.,][]{myers2007clustering, ross2007higher, huterer2013calibration}. We find that all of the mitigation methods are able to reduce such large scale contamination, while there still remains substantial excess clustering on large scales, mostly, with the two traditional linear multivariate methods. The neural network method is much more efficient in reducing such excess. When we investigate the effect of the survey window function on this data, we find that the window effect at $\ell$ < 50 is expected to be less than 5\% (more details presented in Appendix \ref{app:windowfunction}, see Fig. \ref{fig:Cellwindowratio}).\\

In comparison to our default neural network model, we also show the measurements mitigated with the neural network model without the feature selection process labeled as `plain' (blue dashed curves), which is very similar to the default case. In the next section, we test the mitigation methods using the mock datasets for which we know the true clustering signals. As will be demonstrated, our default neural network model with the feature selection process is chosen based on this mock test.

\subsection{Testing the mitigation methods on the mock data}
We treat the mocks as if the contamination model was unknown and apply the mitigation pipeline on the mocks as exactly used for the real dataset. After modeling the selection mask for each mock, we remove the pixels whose selection masks values are < 0.5 or > 2.0. This reduces the mock footprint size from 86,875 to 86,867 pixels. Again, the mocks do not include the redshift-space distortions.

\subsubsection{Feature selection of mock galaxies}
\begin{figure*}
    \centering
    \includegraphics[width=\textwidth]{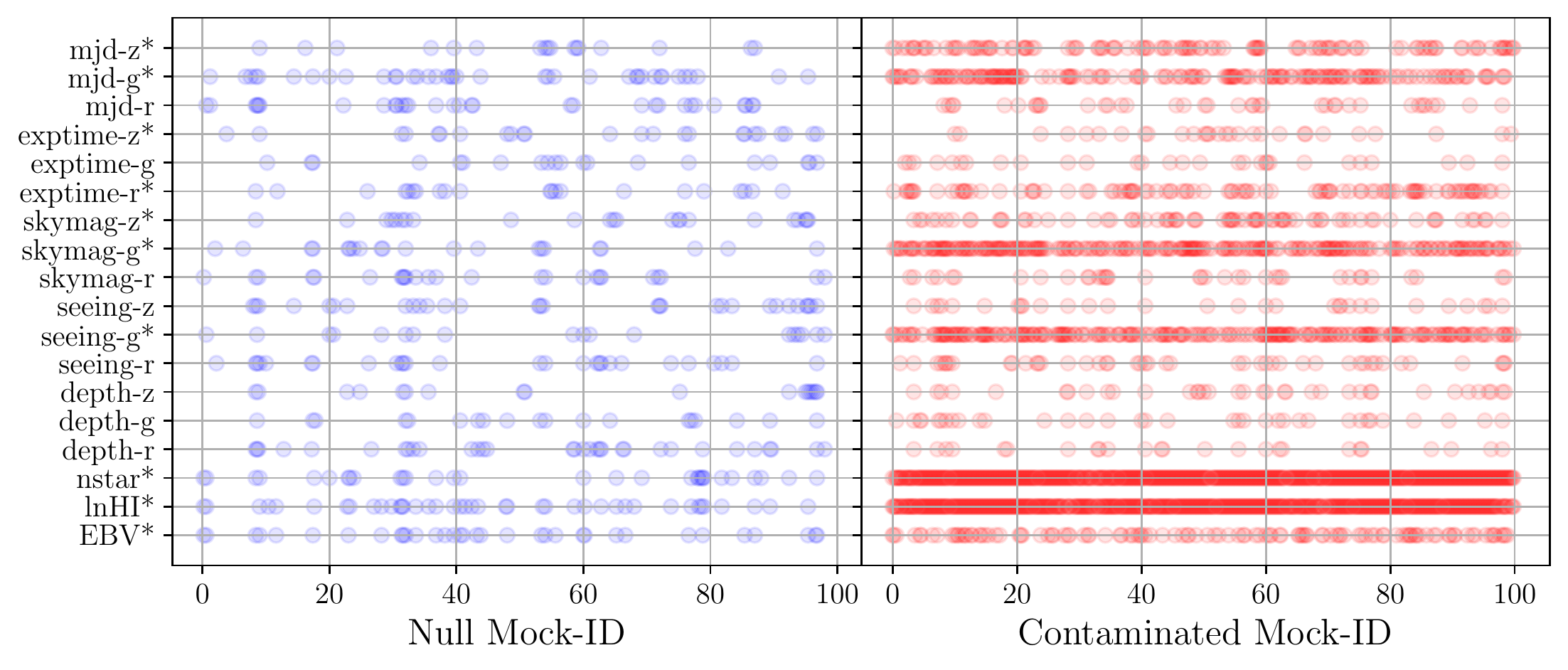}
    \caption{Important imaging maps identified in the mocks by the feature selection procedure for the five partitions of the 100 null (left) and contaminated mocks (right). The maps used in the input contamination model were marked by `*'. The right panel shows that the feature selection procedure has identified EBV, Stellar density, skymag-g, seeing-g as important in most of the contaminated realizations whereas in the left panel, no map is consistently selected as important among all of the 100 null mocks.}
    \label{fig:mockablation}
\end{figure*}

Fig.~\ref{fig:mockablation} shows the distribution of the imaging attributes selected by the feature selection process for all of the five partitions of the 100 null (left) and contaminated (right) mocks. For the null mocks, there is no contamination and the feature selection correctly removes most or all of the imaging attributes, as demonstrated by the sparse distribution of the points in the left panel. The imaging attributes that survived feature selection, probably due to a coincidental correlation with the galaxy density, are randomly distributed. On the other hand, the right panel shows that the feature selection procedure correctly identifies most of the input contamination attributes (marked by `*' on the y-axis) for the contaminated mocks and almost always selects $lnHI$ and $nstar$. Indeed, as shown in Fig.~\ref{fig:dr7ablation}, these two attributes were the two most significant input contamination.

\subsubsection{Mean mock galaxy density}
\begin{figure*}
    \centering
    \includegraphics[width=0.79\textwidth]{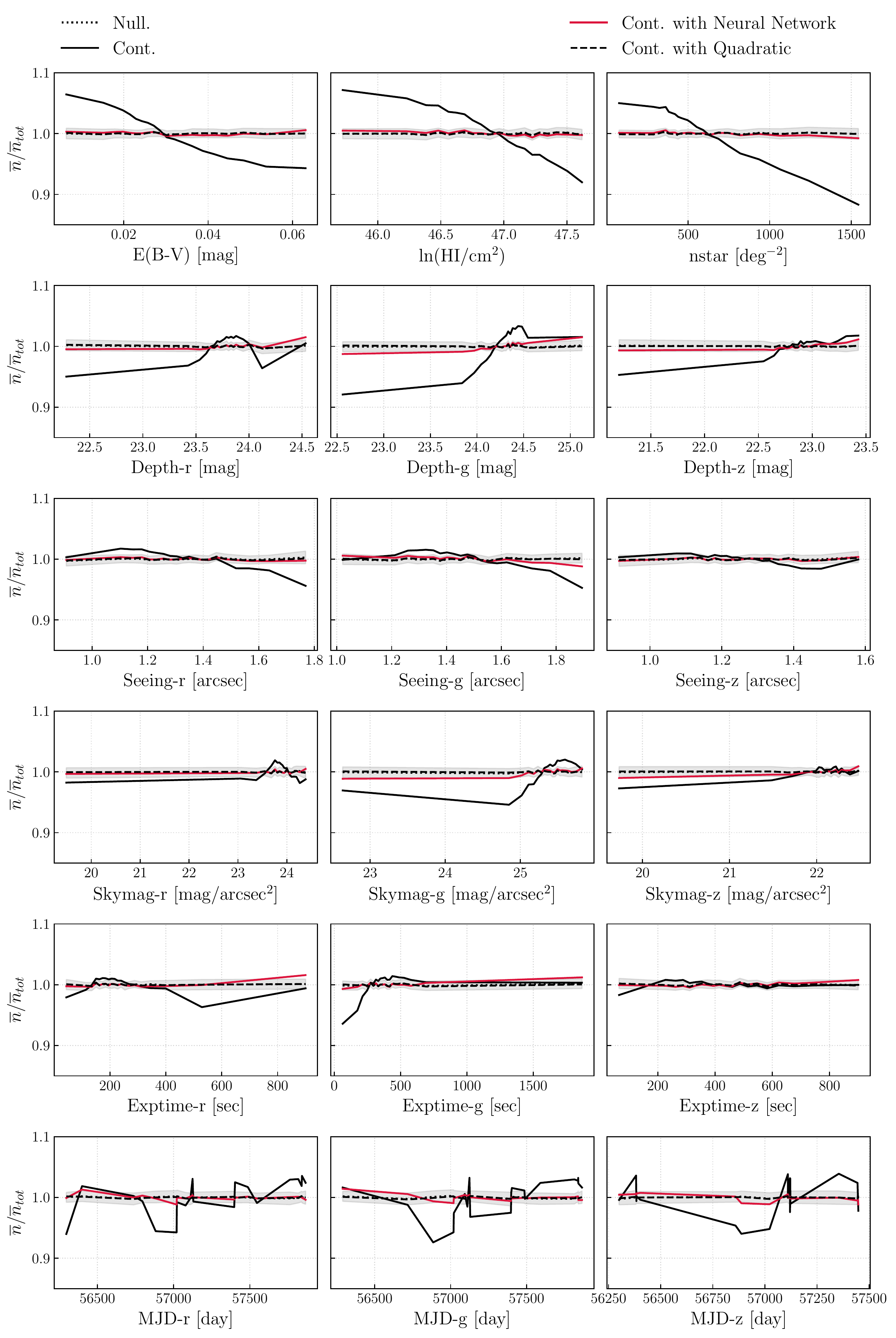}
    \caption{The number density of the mock galaxies as a function of the potential systematics averaged over the 100 mock datasets. The grey shaded region illustrates 1$-\sigma$ dispersion in the null mocks. The dotted curve shows the mean density of the null mocks. The black solid curve shows the mean density dependence on the imaging attributes for the contaminated mocks. The solid red curve shows the mean density after correcting for the systematics with the neural network selection mask. The dashed black curve shows the corrected results with the quadratic polynomial model selection mask. The result of the linear polynomial model is almost unity, and therefore is omitted for clarity.}
    \label{fig:nnbarmock}
\end{figure*}

In Fig.~\ref{fig:nnbarmock} we show the number density of mock galaxies, averaged over the 100 mocks, as a function of the imaging attributes. As expected, the galaxy density of the contaminated mocks shows strong or moderate dependence on $ebv$, $nstar$, $lnHI$, $seeing-g$, $skymag-g$, $skymag-z$, $exptime-r$, $exptime-z$, $mjd-g$, and $mjd-z$ which were indeed the inputs to the contamination model. Meanwhile, the galaxy density also shows strong dependencies on $mjd-r$, $depth-r$, $depth-g$, and $depth-z$ through the correlation between these and the input contamination attributes. Looking at this result alone from a real data perspective, one would not be able to single out the underlying imaging attributes that are directly responsible for the contamination. When the inputs to the mitigation procedure include all of the input contamination maps, Fig.~\ref{fig:nnbarmock} shows that all methods effectively remove the dependence. In subsection \ref{subsec:discussion}, we discuss further how well the underlying true mean density is recovered after mitigation.

\subsubsection{Angular power spectrum of mock galaxies}

\begin{figure*}
\centering
\includegraphics[width=\textwidth]{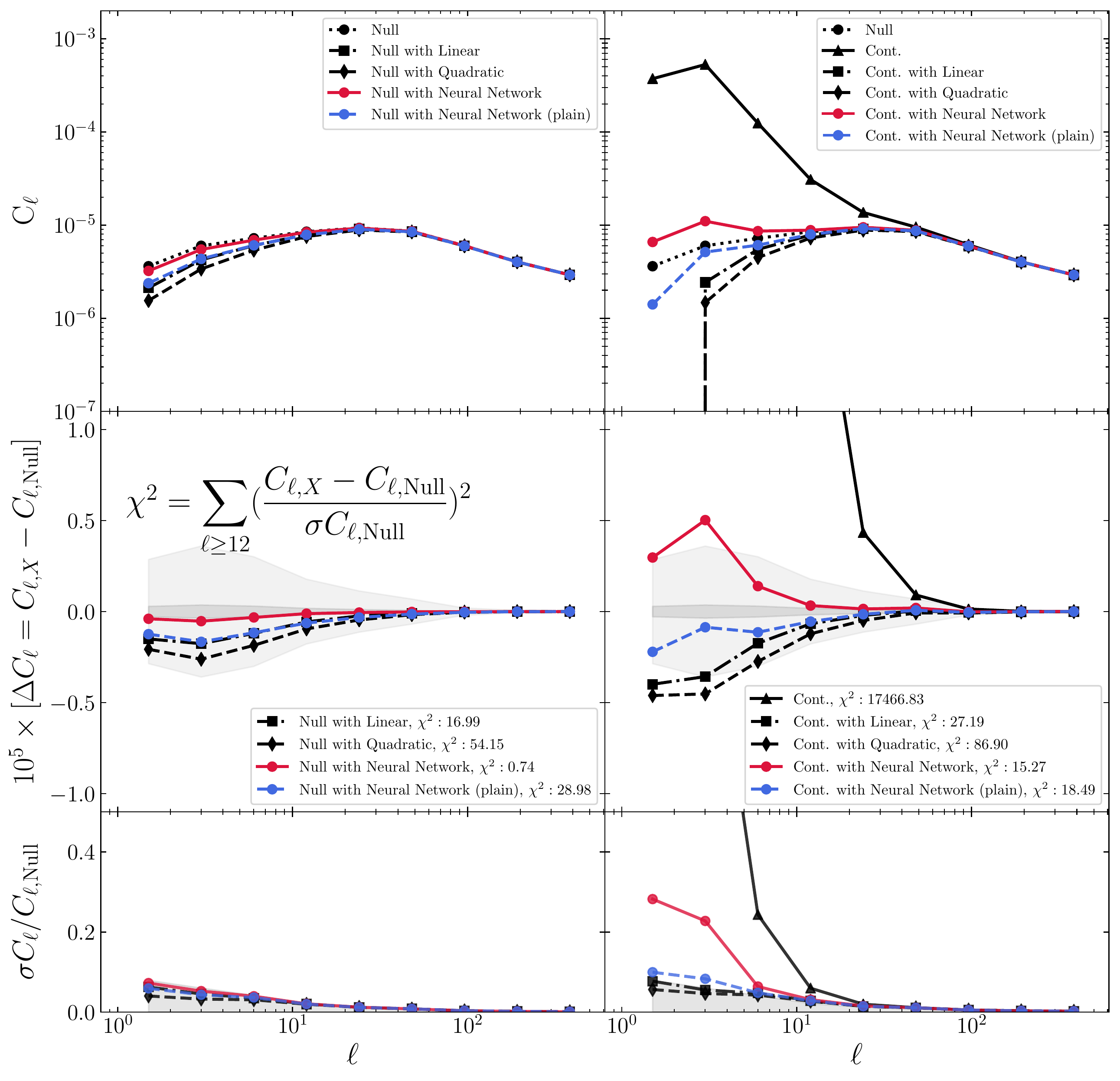}
\caption{\textit{Top row}: The mean angular power spectrum of the 100 contaminated (null) mocks in the right (left) panel. \textit{Middle row}: The mean power spectrum subtracted by the mean of the null mocks to better visualize the remaining bias after each mitigation. The dark grey shaded region shows the 1$-\sigma$ confidence region of the mean of 100 mocks, while the light grey area shows the 1$-\sigma$ confidence region for one mock, calculated from the dispersion of 100 mocks. To account for the increased shot noise during contamination, we remove the same constant power from all contaminated/mitigated power spectra until their small scale power matches that of the null mock power spectrum. We quantify the significance of the remaining bias by calculating $\chi^2$, the sum of the squared offset weighted with the diagonal variance of the mean $C_\ell$ of null mocks over the last six bins ($\ell~\geq 12$). The middle panel on the left illustrates the neural network without feature selection (`plain') tends to remove the cosmological clustering signal. \label{fig:deltaclmock}}
\end{figure*}

Fig. \ref{fig:deltaclmock} shows the mean angular power spectrum of the 100 null and 100 contaminated mocks in the left and right panel, respectively, in the top row. In the middle row, we illustrate the remaining bias\footnote{Due to the two-step noise introduced during contamination, the shot noise of the contaminated power spectra is increased by almost a factor of two. We estimate the total offset noise to be around $3.05\times10^{-6}$ from $\sigma^2(ngal)/\overline{n}^2$ and subtract it from the power spectrum of the contaminated mocks. The additional shot noise is mostly originated from the Poisson process we applied, i.e., precisely 1/$\overline{n}$, where $\overline{n}$ is the average number density after contamination, while an extra $\sim$ 10\% is also due to the noise we added to the contamination model.} on clustering after mitigation as an offset from the true power spectrum (i.e., the null power spectrum from the left panel\footnote{Note that we use the clustering of the null mocks before mitigation as the ground truth model since the survey window for all of the mocks before and after systematics treatment does not change.}). One can see that the contamination substantially increased power at $\ell < 50$. Since the contamination model is based on the linear polynomial model, all three fitting methods, i.e., the linear, quadratic, and neural network, are capable of reproducing the true input contamination, while they perform differently in the presence of the two layers of noise we added and the eight additional imaging attributes that are non-trivially correlated with the ten input contamination attributes. \\

In the right top and middle panel, we find all three methods effectively remove the contamination over $\ell < 100$; in detail, the linear (black dot-dashed) and the quadratic polynomial (black dashed) methods slightly over-correct power while the neural network method (solid red) slightly under-corrects it. Note that, without the feature selection process (dashed blue), the neural network method would also over-correct the large scale power like the linear and quadratic polynomial models. The left top and middle panel show that, in the absence of contamination, both the linear and quadratic polynomial methods over-correct the large scale power since the fitting methods can always find purely coincidental consistency between the imaging attributes and the cosmic variance. The quadratic method that has a greater freedom appears more prone to such problem. On the other hand, the neural network method without the feature selection process shows a lesser degree of overfitting than the linear methods, probably due to the validation procedure. Our default neural network method, which incorporates feature selection, is the most robust mitigation methodology against overfitting.\\

\begin{figure*}
\centering
\includegraphics[width=\textwidth]{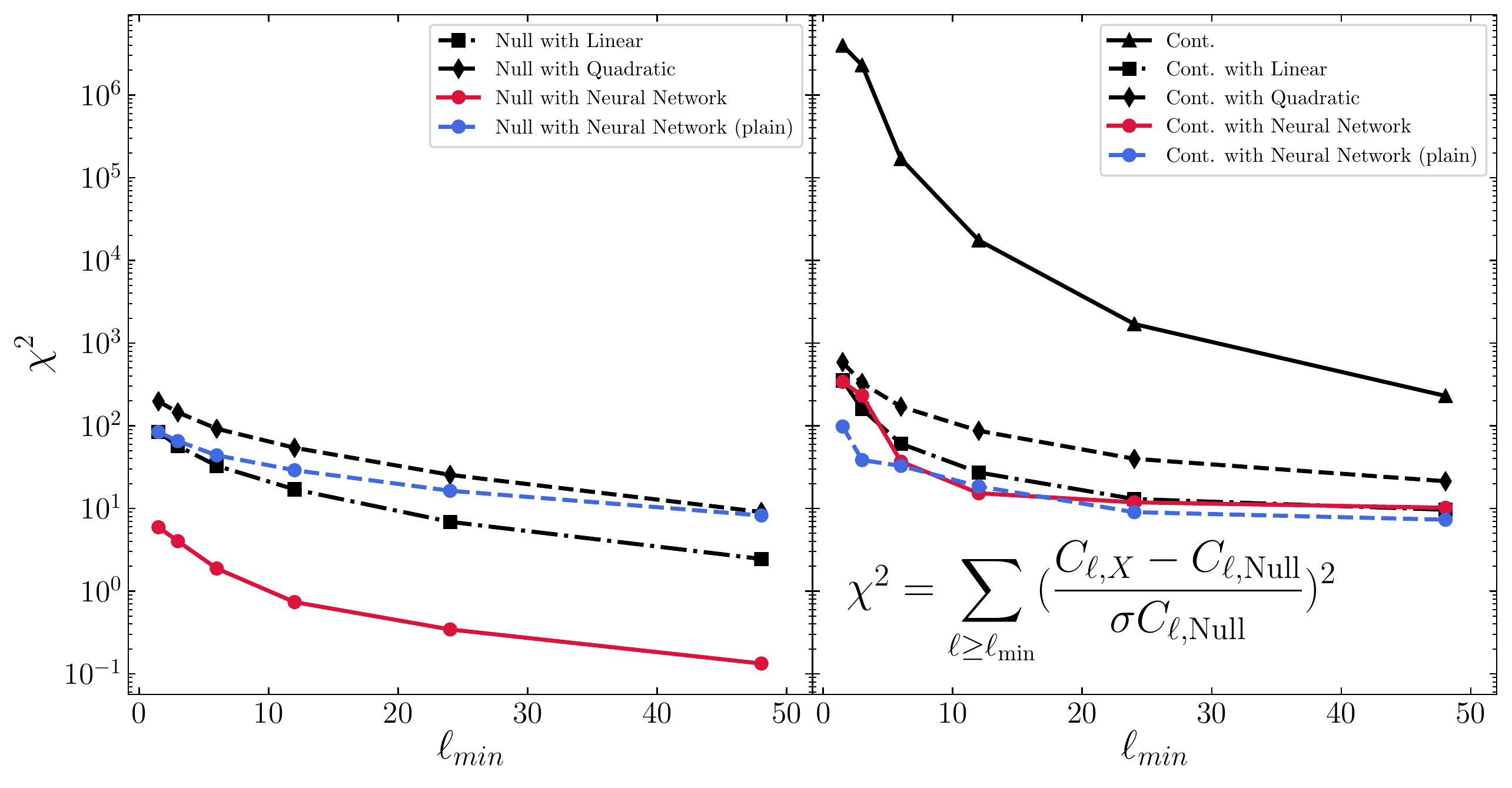}
\caption{Dependence of $\chi^{2}$ on the lowest bin $\ell_{\rm min}$ in the null (left) and contaminated (right) mocks. To better quantify the residual bias introduced by each method, we evaluate the dependence of the bias on the lowest bin $\ell_{{\rm min}}$ that is included in the $\chi^{2}$ computation. The default neural network method performs significantly better than the conventional methods for the null mocks, mainly because the feature selection procedure successfully prevents the method from regressing out the cosmological clustering signal. For the contaminated mocks, all methods tend to perform similarly, as expected, since all mitigation methods can reproduce the input contamination model. }\label{fig:chi2clmock}
\end{figure*}

The remaining bias can be compared to the typical error expected for such data. The dark and light grey shaded regions in the middle panels indicate the 1-$\sigma$ confidence regions for the mean and the individual mock of the 100 mocks, respectively.  We quantify the significance of such remaining bias by calculating $\chi^2$, the sum of the squared offset weighted with the diagonal variance for the mean at each $\ell$ bin. Note that we use the variance from the 100 null mocks for calculating $\chi^2$ of the contaminated mocks in order to avoid an advantage of the increased variance after contamination. We find that the default neural network with $\chi^{2}=0.74$ (reduced $\chi^{2} = 0.12 $ with $dof=6$) recovers the true underlying clustering when applied to the null mocks well within $1\sigma$ C.L. of the sample variance. We estimate the significance with taking the residual systematics as one extra degree of freedom,
\begin{equation}\label{eq:chi2sigma}
 \text{max systematics} \sim \sqrt{\chi^{2}}
\end{equation}
On the other hand, the linear and quadratic models have systematic biases with more than $4\sigma$ and $7.4\sigma$ significance. When applied to the contaminated mocks, we find that the neural network method returns$\chi^2=15.3$ from six $\ell$ bins ($\ell \geq 12$), while the linear and quadratic methods, respectively, polynomial mitigation return $27.2$ and $86.9$. The difference in $\chi^2$ among the different mitigation methods is not significant compared to the $\chi^2=17,466.8$ before mitigation. While the neural network model appears to perform the best, $\chi2$ of $15.3$ (reduced $\chi^2$ of $2.6$) indicates that the residual is much more significant than the sample variance. However, for the contaminated mocks, we added substantial statistical noises, almost doubling the noise. If we use the covariance of each contaminated/mitigated case, we get $\chi^2$ of $7.7$ (reduced $\chi2$ of $1.3$) for our default case; that is our residual is at the level of the statistical noise we added in the process of contamination. These $\chi^{2}$ values can be translated to $2.8-3.9\sigma$ uncertainties in the residual systematics (see Eq. \ref{eq:chi2sigma}) which implies any particular cosmological study requires a more thorough analysis of systematics to determine an estimate of the residual systematic uncertainty for the parameters of interest.\\

Such $\chi^2$ can depend on the lower limit of $\ell$ we consider. In Fig. \ref{fig:chi2clmock}, we investigate the behavior of the remaining bias depending on $\ell_{min}$, which shows that the neural network method consistently returns a lower remaining bias for various $\ell_{\rm min}$ choices. However, for the contaminated mocks, the difference is small and we conclude that all mitigation methods perform similarly for the contaminated mocks.\\

The bottom panels of Fig.~\ref{fig:deltaclmock} compare the noise introduced by the three different mitigation processes. The right panel shows that the contamination process (solid black) introduces additional noise on large scales relative to the variance of the null mocks (the gray shade). After mitigation, the variance is reduced, which is probably related to the decrease in the large scale power, since the variance of power is proportional to the amplitude of power itself in the Gaussian limit. The quadratic method shows the smallest fractional error on large scales due to the reduced amplitude after correction. In all cases, if we calculate the fractional variance with respect to the measured $C_\ell$ (e.g., $\sigma C_{\ell, NN}/C_{\ell, NN}$ instead of  $\sigma C_{\ell, NN}/C_{\ell, Null}$), it agrees with the fractional variance of the null mock (gray shade). Therefore, we do not observe a nontrivial increase in variance by any of the mitigation methods we tested.

\subsubsection{Cross power spectrum of mock galaxies and imaging attributes}
\begin{figure*}
\centering
\includegraphics[width=0.78\textwidth]{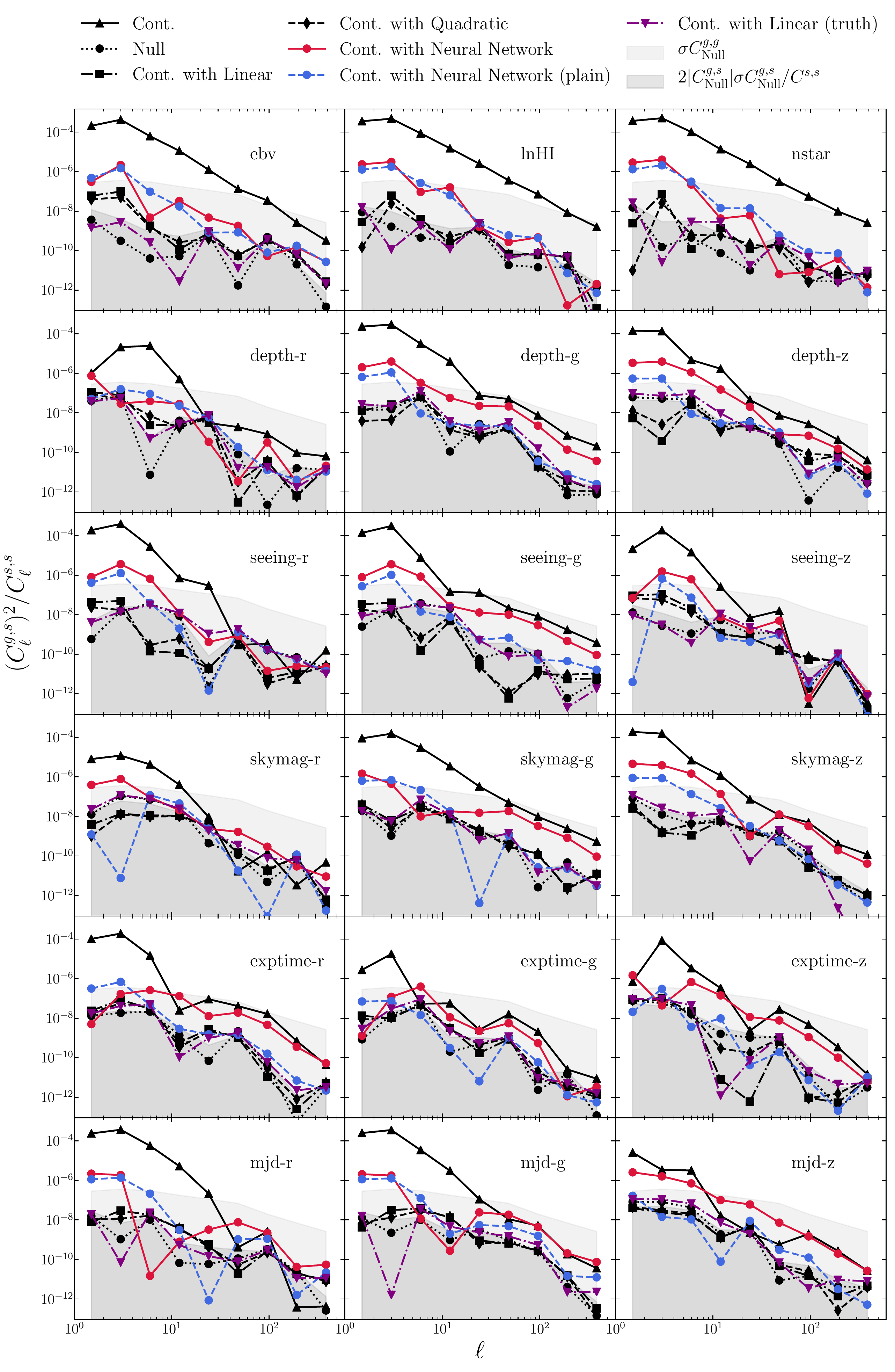}
\caption{The mean cross power spectrum of the contaminated mock catalogs and the imaging attributes for different mitigation techniques. Our default neural network method with feature selection is shown in solid red while the performance without feature selection (`Neural Network (Plain)') is shown in dashed blue. The dark grey shaded region shows the 1$\sigma$ confidence region of the plotted quantity derived from $2\sigma(C_{g,s})*C_{g,s}/C_{s,s}$, and the light grey region shows the typical 1-$\sigma$ confidence region of the mean auto power spectrum of the 100 mocks. The mitigation with the ground truth contamination model is shown with a purple dot-dashed curve as `cont. with linear (truth)'. \label{fig:clcrossmock}}
\end{figure*}

In Fig. \ref{fig:clcrossmock}, we show the mean cross power spectrum of the 100 mock catalogs and the imaging attributes for the different mitigation techniques. All three methods substantially reduce the cross power with the imaging attributes. The neural network method tends to show a small residual for $\ell <10$ that is greater than those of the linear and quadratic polynomial models. The dark shaded region shows the 1$-\sigma$ confidence region of the mean cross power propagated to $C_{s,g}^{2}/C_{s,s}$ and the light shaded region shows the 1$\sigma$ confidence interval of the mean auto-power spectrum of the mocks as shown in Fig. \ref{fig:clcross}. Therefore, we find that these residuals are greater than the statistical noise of $C_{s,g}^{2}/C_{s,s}$, but  the effect on the \textit{auto power spectrum} are marginal for $\ell~ >~ 10$. This excess on small $\ell$ is partly due to the greater auto galaxy power spectrum amplitude (in Fig.~\ref{fig:deltaclmock}) after mitigation than those by the other methods. Meanwhile we still find a residual correlation with skymag-z and mjd-z even after accounting for the effect of the auto power spectrum amplitude. Without the feature selection procedure (dashed blue), the neural network also returns a smaller residual. In essence, we see that once feature selection is applied, the neural network only corrects to a certain level, controlled by the specifics of the feature selection procedure. This protects against over-fitting due to random correlations between the imaging attributes and the galaxy density field.\\

As a sanity check, if we use the true input linear contamination model to mitigate the systematics (purple dot-dashed line in Fig. \ref{fig:clcrossmock}), the cross-correlation completely vanishes as expected\footnote{The auto-power spectrum of the contaminated mocks mitigated with the true input contamination model (in Fig. \ref{fig:mockdclextra}) returns the smallest residual bias relative to the uncontaminated clustering, as expected.}. For the null mocks, all mitigation methods return negligible cross-correlation, which is omitted from Fig. \ref{fig:clcrossmock} for clarity.\\

Overall, while the cross-correlation statistics between the galaxy density and the systematics attributes are a useful indicator for the level of contamination, we find it may be difficult to infer and discriminate the level of contamination in the density field from such cross-correlation statistics beyond what can be probed by the auto power spectrum.

\subsubsection{A case with underfitting}
\begin{figure*}
\centering
\includegraphics[width=\textwidth]{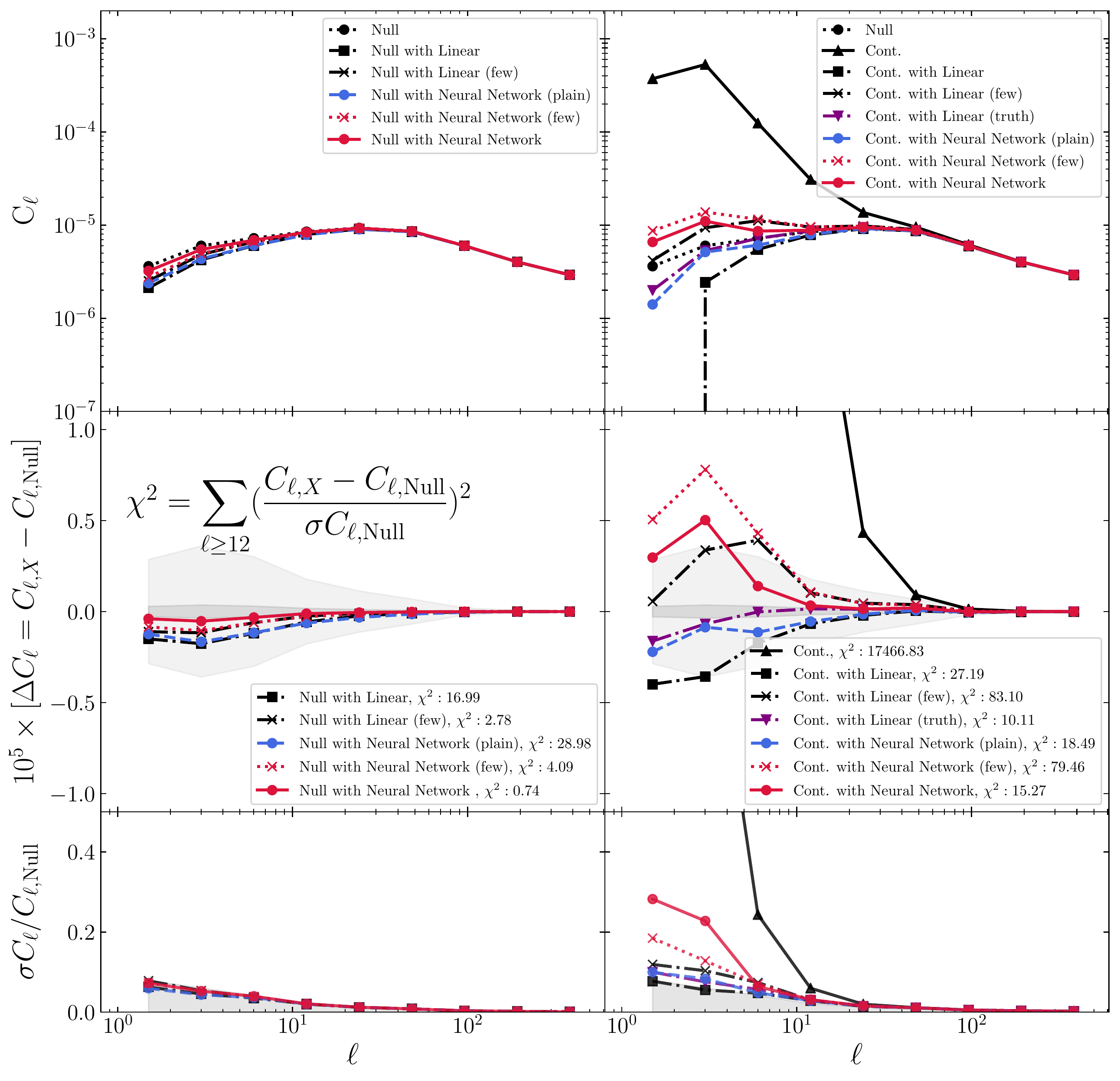}
\caption{Same as Fig. \ref{fig:deltaclmock} showing the mean auto power spectrum of the 100 null (left) and contaminated (right) mocks mitigated with the fewer imaging maps `few' (to demonstrate under-correction), neural network without the feature selection `plain', and the ground truth contamination model `truth'. The mitigation with the ground truth model achieves the lowest residual bias as expected. For the null mocks, using fewer imaging maps prevents over-fitting simply by providing less freedom in the regression model while it leads to underfitting for the contaminated mocks. }
\label{fig:mockdclextra}
\end{figure*}

It is possible that we may identify only a subset of the contamination attributes for a given data set and attempt to mitigate the contamination based on such limited information. We consider a situation where we input only five imaging attributes to the mitigation procedure: four from the true contamination inputs, i.e., $EBV$, $lnHI$, $nstar$, $skymag-g$ and one that is not among the true contamination inputs, but correlated with the contamination inputs, i.e., $depth-r$. The neural network method could be more resilient to such limited information since its nonlinear activation function may allow the mitigation procedure to better utilize the correlation between different input imaging attributes. In Fig.~\ref{fig:mockdclextra}, the linear polynomial method with `few' inputs shows a lesser degree of overfitting for the null mocks, compared to the default linear case, while showing under-fitting for the contaminated mocks. This is expected as the freedom of the linear model is now limited.  The neural network method with the `few' inputs (without feature selection) returns a very similar pattern as the linear `few' case, which implies that our current default neural network method does not have an advantage over the linear model in such a case despite its greater flexibility. This underfitting case may be worth future investigation, as apparent excess clustering remains in DR7 in Fig.~\ref{fig:clxi} when any method is applied, but especially in the case of the application of multivariate linear models.

\subsection{Summary and Discussion}\label{subsec:discussion}
In summary, we find that our default neural network method is more robust against overfitting based on the test with the null mocks. This is due to the feature selection process that appropriately reduces the flexibility of mitigation. Based on the tests with the contaminated mocks, we find that both the linear and neural network methods perform equally well in terms of the residual bias, while the neural network method is more robust against overfitting. The quadratic polynomial method appears to be more prone to the overfitting problem than the other two methods since it has a greater flexibility than the input contamination model, but without a way to suppress the flexibility. All methods do not increase fractional variance during the mitigation process. Note again that we deliberately choose the linear model in contaminating the mocks in this test in order to prevent a disadvantage in using the linear and quadratic polynomial methods.  Therefore, the decent performance of the linear method is warranted. In real data, the contamination due to observational effects can take a more complex form as implied by the difference in the mitigation results between the data and our mocks. Therefore our mock test is a conservative estimation of the comparative mitigation capability of our default neural network method. \\

While we demonstrated qualitatively and quantitatively that our fiducial neural network method is more robust than the conventional methods, for both DR7 as well as for the mocks, Fig. \ref{fig:nnbar}-\ref{fig:clcross} show non-negligible residual contamination compared to the expectations. It is not surprising that the mitigation of imaging systematic effects in the real data is more challenging than that of the linear-model based systematics in our mock tests. In this subsection we attempt to provide a quantitative evaluation of the residual systematics of DR7.\\

\begin{figure}
    \centering
    \includegraphics[width=0.45\textwidth]{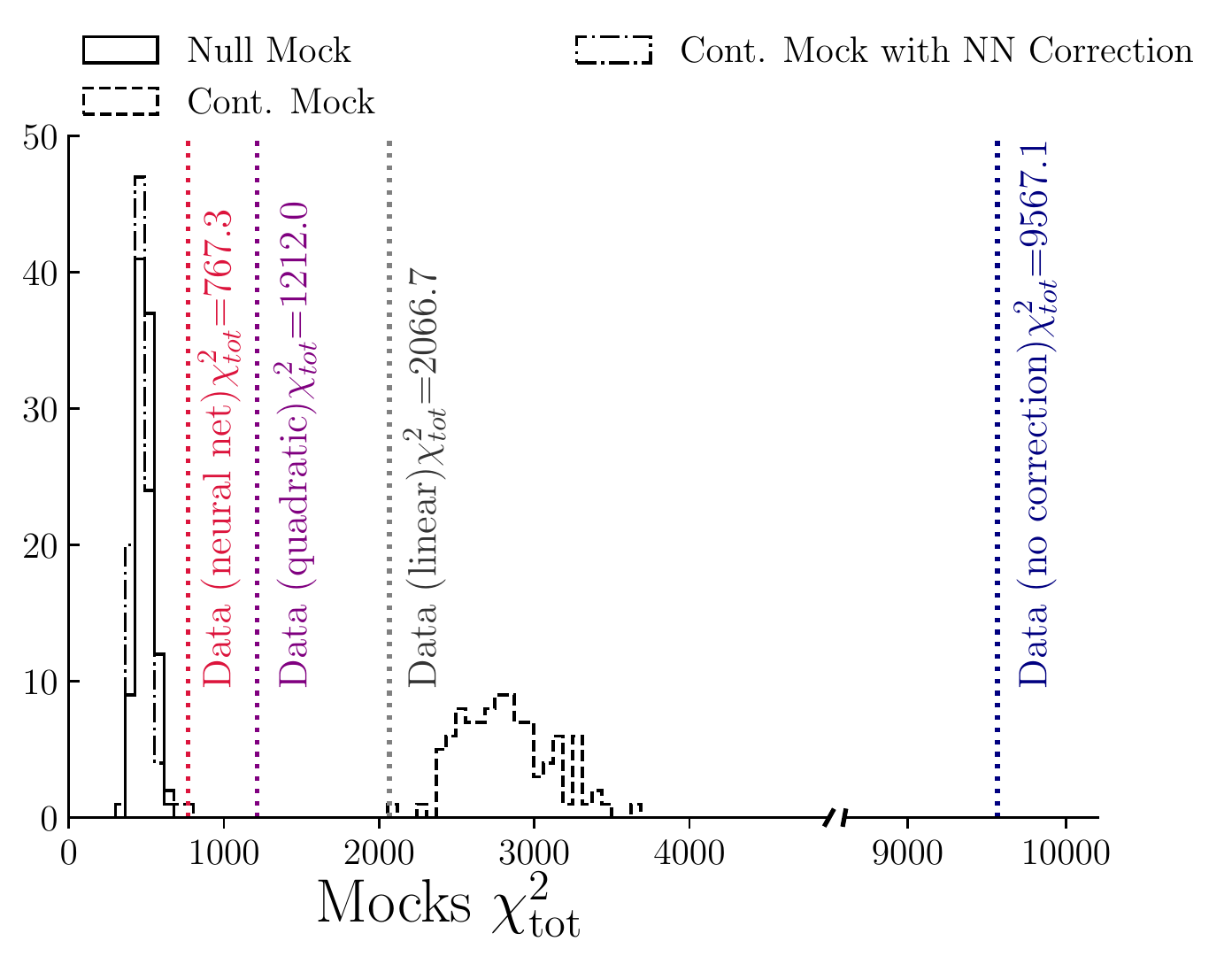}
    \caption{\textit{Left}: $\chi^{2}$ distribution for the null mocks (solid black), contaminated mocks (dashed black), and contaminated mocks with NN mitigation (dot-dashed black). The vertical dotted lines overlay the $\chi^{2}$ values for the data with the default NN (red), quadratic (purple), and linear (grey) treatments. The $\chi^{2}$ statistics before treatment is shown on the right (dark blue).}
    \label{fig:chi2pdf}
\end{figure}

We quantify the residual systematics in the mean density against the 18 imaging maps using $\chi^2$ of the mean density diagnostic as in Fig. \ref{fig:nnbar} and Table \ref{tab:chi2}~\footnote{To compare these with the mock results, we limit DR7 to the mock footprint, and therefore the $\chi^2$ results are slightly different from Fig. \ref{fig:nnbar} and Table \ref{tab:chi2}. The mock footprint is smaller than the data footprint by almost a factor of two.}. The null hypothesis is that the total residual squared error of the mean density observed in the data should be consistent with the distribution of the $\chi^{2}$ statistics constructed with the null mocks. The vertical lines in Fig. \ref{fig:chi2pdf} present the $\chi^{2}$ values observed in the data for before systematics treatment (9567.1) and after treatment with linear (2066.7), quadratic (1212.0), default Neural Network (767.3), and  Neural Network plain (744.4) approaches. As a comparison, we present the distributions of the $\chi^{2}$ observed in the null mocks (solid), contaminated mocks (dashed), and contaminated mocks after neural network mitigation (dot-dashed). Fig. \ref{fig:chi2pdf} illustrates a factor of 12 improvement in terms of the residual $\chi^2$ when using the neural network. Compared to the conventional quadratic method ($\chi^2=1212$), the NN-based method makes a factor of 1.6 improvement.\\

The mean and standard deviation for the distribution of $\chi^{2}$  values observed in the null mocks are $487.1 \pm 4.92$, respectively. The same quantities observed in the contaminated mocks are $2819.0 \pm 29.30$, while Neural Network mitigation decreases these statistics to the mean of $472.0\pm 6.59$. We perform Welch's t-test \citep{welch1947generalization} on the $\chi^{2}$ distributions of the null mocks and the mitigated contaminated mocks, and conclude that the two distributions have identical means with $t-statistic=1.8$, $p-value=0.06$ and the neural network recovers the underlying cosmological mean target density. Note that the mean $\chi^{2}$ observed in the contaminated mocks, 2819.0, is much smaller than the value observed in the data (the blue vertical line, 9567.1). This indicates that the systematic effects rigorously simulated in the mocks are still not as strong as the systematics in the real data or the effect of the non-diagonal terms, that we ignored in the covariance, is different in the data and in the mocks.\\

We perform a hypothesis testing given the $\chi^{2}$ value observed in the data after the neural network mitigation and the distribution of $\chi^{2}$ values observed in the 100 null mocks.
If the imaging maps were independent and normal deviates, the $\chi^{2}$ values must follow a Chi-squared distribution for $dof=360$ where 360 is the total number of bins. This assumption is not necessarily satisfied given the correlation among imaging systematics, and we indeed find $\chi^{2}$ of the null mocks is better fit with $dof=487$. Therefore, we assume that the underlying $dof$ for evaluating $\chi^2$ is 487. For DR7, the $\chi^2$ decreased from 9567 before contamination to 767.3 after our NN-based mitigation. Compared to $dof=487$ we expect for the approximate truth, this is a substantial excess, being very unlikely due to random fluctuation: for example, the top 5\% of distribution with $dof=487$ is at $\chi^2 = 539.4$, which is 40\% lower than 767.3. Assuming a mere random fluctuation is bounded at this upper 5\% and the $\chi^{2}$ values of the data are consistent with that of the mocks, this offset could imply that we underestimated the covariance at least by $\sqrt{(767.3/539.4)} = 19\%$ or that our model, when applied to DR7, missed at least 19\% of the systematic effects in the target density.\\

\begin{figure}
    \centering
    \includegraphics[width=0.45\textwidth]{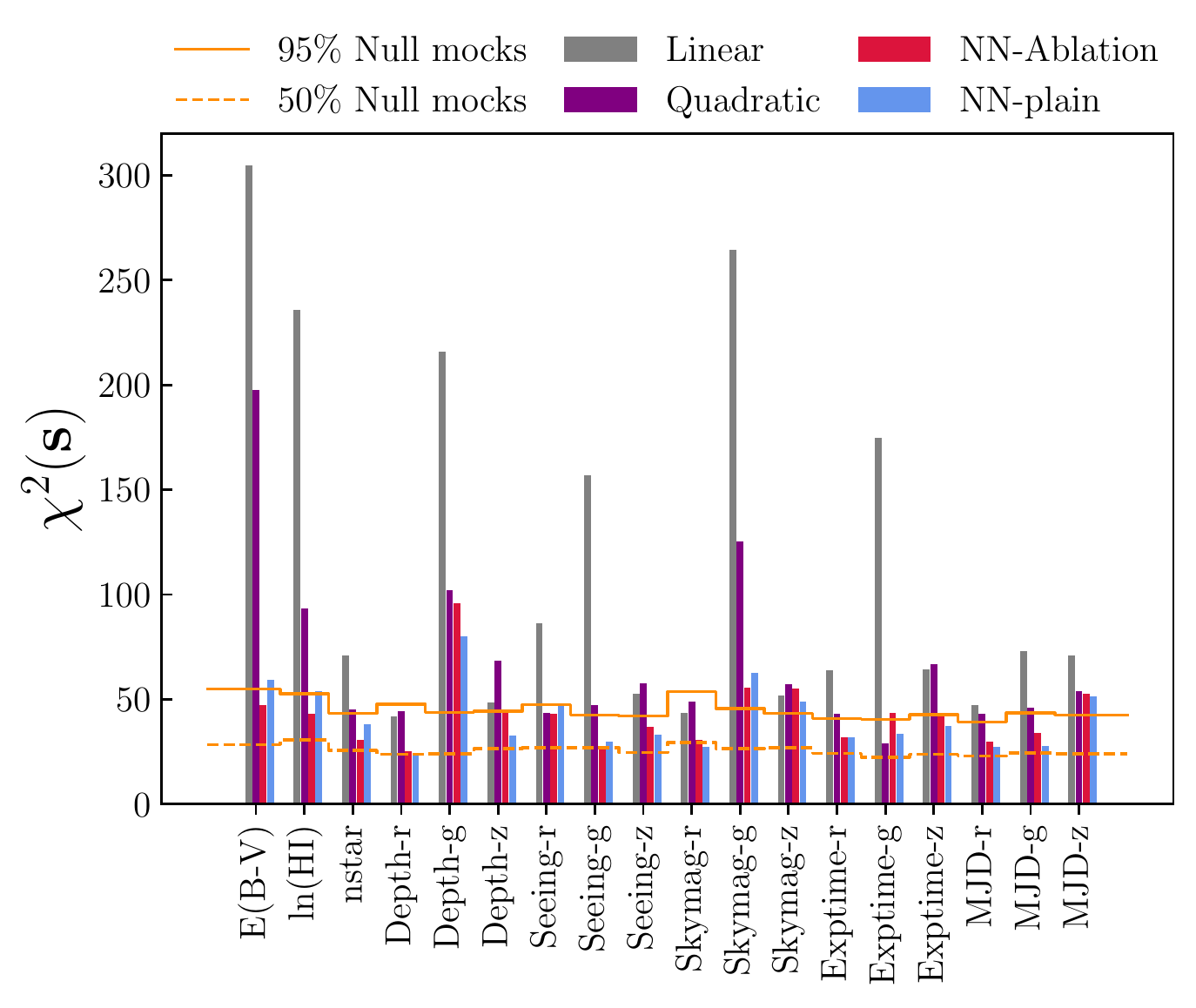}
    \caption{The breakdown of $\chi^{2}$ values observed in the data on the mock footprint after linear (grey), quadratic (purple), neural network with feature selection (red), and neural network plain (blue) treatments. We also plot the 50- and 95-th percentiles of the same quantity observed in the null mocks in orange curves (note that $\chi^{2}(\textbf{s)}$ is not a continuous quantity). Given the 5\% threshold, we can argue that there exists known residual systematics against E(B-V), depth-g, skymag-g, skymag-z, and MJD-z.}
    \label{fig:chi2breakdown}
\end{figure}

We further investigate the contribution of each imaging map to the residual systematics in DR7. Fig. \ref{fig:chi2breakdown} presents the $\chi^{2}$ vs each imaging map after the linear (grey), quadratic (purple), neural network with feature selection (red), and neural network plain (blue) treatments. The statistics before mitigation are not shown for clarity. We also plot the 50- and 95-th percentiles of the same quantity observed in the null mocks in orange horizontal curves. Depth-g seems to be the main source of the residual systematics which is not mitigated even by the neural network-based methods. All methods also show residual systematics against skymag-g, skymag-z, and MJD-z. The standard treatments show some additional residual systematics against E(B-V), lnHI, and exptime-z. We perform further tests by masking out high extinction and low depth-g regions. We find that applying a more rigorous cut on depth-g (e.g., depth-g $> 24.95$) yields a more cleaner mean density at the expense of losing 9\% of the data. Although our objective was to avoid applying rigorous cuts on imaging and demonstrate the gain by using non-linear methods, our tests suggest that careful masking based on imaging, particularly depth-g, seems necessary for cosmological studies. \\

In summary, our results indicate that any cosmological study requires a thorough analysis of systematics to determine an estimate of the residual systematic uncertainty for the parameters of interest. In order to use this sample for a cosmological exploitation, we would need to account for 19\%  (or more, depending on the assumption on the baseline) additional systematic errors to the statistical errors in the density field level. Our analysis also suggests that additional masking, especially based on depth-g, or improving the method to deal with depth and sky background issues would be the next steps to prepare this data for cosmological analysis. We leave this for future study, as the focus of this study is to compare our non-linear, neural network method to other map-based methods.

%% file: sections/conclusion.tex
\section{conclusion}\label{sec:conclusion}
In this paper, we have presented a rigorous application of an artificial neural network methodology to the mitigation of the observational systematics in galaxy clustering measurements of an eBOSS-like ELG sample selected from DR7 (see \S~\ref{sec:data}). We have investigated the galaxy density dependency on 18 imaging attributes of the data (see Fig. \ref{fig:eboss_dr7}). We compare the performance of the neural network with that of the traditional, linear and quadratic multivariate regression methods. The key aspects of our neural network methodology are:\\

\begin{itemize}
    \item The application of k-fold cross-validation, which implements the training-validation-test split to tune the hyper parameters by evaluating how well the trained network generalizes to the unseen,  validation data set and therefore to suppress overfitting when applied to the test set;
    
    \item The repeated split process until we cover the entire data footprint as test sets;
    
    \item The elimination of redundant imaging maps by the feature selection procedure to further reduce the overfitting problem and therefore protect the cosmological clustering signal.
\end{itemize}

We apply the output of our pipeline, i.e., the selection mask for the DR7 footprint to the observed galaxy density field. Benchmark selection masks are also produced employing the linear and quadratic polynomial regression. Comparing statistical results before and after applying the selection masks, we find that:\\
\begin{itemize}
    \item Galactic foregrounds are the most dominant source of contamination in this imaging dataset (see Figs. \ref{fig:nnbar}, \ref{fig:clcross}, and \ref{fig:xicross}).
    
    \item This contamination causes an excess clustering signal in the auto power spectrum and correlation function of the galaxy density field on large scales (see Fig. \ref{fig:clxi}).
    
    \item All mitigation techniques e.g., the neural network method as well as the linear multivariate models using the linear and quadratic polynomial functions, are able to reduce the auto and cross clustering signals (see Figs. \ref{fig:xicross} and \ref{fig:clcross});
    
    \item However, the neural network removes the excess clustering more effectively in the auto power spectrum and correlation function of galaxies (see Fig. \ref{fig:clxi}).
\end{itemize}

The last result implies that our neural network method has a higher flexibility than both linear multivariate models we tested, and it is therefore capable of capturing the non-linear systematic effects in the observed galaxy density field.\\

We apply our methodology on two sets of 100 log-normal mock datasets with (`contaminated mocks') and without (`null mocks') imaging contamination to evaluate how well the ground truth cosmological clustering can be reconstructed in both cases, and therefore to validate the systematic mitigation techniques. All mitigation techniques are applied in the same way we treat the real data. The key results of our mock test are as follows:\\

\begin{itemize}
    \item The feature selection procedure is able to identify most of the ten contamination input maps as important for the contaminated mocks while correctly identifying most of the maps as redundant for the null mocks (see Fig. \ref{fig:mockablation}).
    
    \item All three mitigation methods, i.e., the linear polynomial, quadratic polynomial, and neural network methods, perform similarly in terms of the residual bias in the presence of contamination. This is expected since the contamination model is based on the linear polynomial model which all three methods are capable of reproducing. The default neural network tends to slightly under-correct which is the outcome of the feature selection procedure. On the other hand, the linear and quadratic polynomial methods tend to slightly over-correct (see the right panel of Fig. \ref{fig:deltaclmock}).;
    
    \item In the absence of contamination, the neural network is the most robust against regressing out the cosmological clustering. This is mainly due to the feature selection process that appropriately reduces the flexibility of the mitigation (see the left panel of Fig. \ref{fig:deltaclmock}). Based on this result, we implement the feature selection procedure for DR7.
    
    \item Using $\chi^{2}$ statistics, we quantify the bias and find that for the null mocks, the default neural network recovers the underlying clustering within $1\sigma$ C.L. (see Eq. \ref{eq:chi2sigma}) while the other methods return more than $4\sigma$ C.L. bias. For the contaminated mocks, all of the methods return biased clustering with $2.8-3.9\sigma$ C.L. which indicates that it is crucial for cosmological parameter estimation to determine the residual systematic uncertainty in scales sensitive to the parameters of interest (see the middle panels of Figs \ref{fig:deltaclmock} and \ref{fig:mockdclextra}).
    
    \item All methods do not increase fractional variance during the mitigation process (see the bottom row of Fig. \ref{fig:deltaclmock}).\\
        
\end{itemize}

We also employ the mocks to investigate the remaining systematic effects in the data (Figs \ref{fig:nnbar}-\ref{fig:xicross}). While the neural network methods outperform the conventional methods, we conclude that the data exhibit around 19\% residual systematics in the target number density (Figs \ref{fig:chi2pdf} \& \ref{fig:chi2breakdown}). Our analysis suggests that a more rigorous masking on  $depth-g$ (e.g., $depth-g>24.95$) improves the mean density at the cost of losing $9\%$ of data. To use this sample for cosmology, we therefore suggest a) accounting for 19\%  (or more, depending on the assumption on the baseline) additional systematic errors to the statistical errors in the density field level; b) performing further analysis of systematics and improvement of the mitigation method to deal with the depth and sky background issues.\\

To conclude, our analyses illustrate that the neural network method we developed in this paper is a promising tool for the mitigation of the large-scale spurious clustering that is likely raised by the imaging systematics. Our method is more robust against regressing out the cosmological clustering than the traditional, linear multivariate regression methods. Such improvement will be particularly crucial for an accurate measurement of non-Gaussianity from the large-scale clustering of current eBOSS and upcoming DESI and the LSST surveys. Our method is computationally less intensive than other approaches such as the Monte Carlo injection of fake galaxies: analyzing DR7 using our default neural network method requires less than six CPU hours. Application of our methodology on any imaging dataset would be straightforward. Our systematics mitigation methodology pipeline is publicly available at \url{https://github.com/mehdirezaie/SYSNet}.

%% file: sections/acknowledgement.tex
\section*{Acknowledgements}
We would like to thank the anonymous referee for constructive comments. M.R. and H.-J.S.~are supported by the U.S.~Department of Energy, Office of Science, Office of High Energy Physics under Award Number DE-SC0014329. This research used the Dark Energy Spectroscopic Instrument (DESI) allocation resources of the National Energy Research Scientific Computing Center (NERSC), a U.S. Department of Energy Office of Science User Facility operated under Contract No. DE-AC02-05CH11231. M.R. would like to thank Stephen Bailey, NERSC, and the DESI collaboration for providing computing resources to carry out the initial phases of this work; Anand Raichoor for providing the stellar mask and discussions on targetting; Nick Hand and Yu Feng for helping with Nbodykit and mock generations; Jeffrey Newman, Arnaud De-Mattia, and Marc Manera for discussions on machine learning and cross-validation; Ted Kisner, Rollin Thomas, and Joel Brownstein for helping with high-performance computing; Dustin Lang, John Moustakas, and Arjun Dey for discussions about the Legacy Surveys and data reduction pipelines. M.R. and H.J. would like to thank Patrick McDonald for discussions on window effects and error analyses. AJR is grateful for support from the Ohio State University Center for Cosmology and Particle Physics. We would like to appreciate the open-source software and modules that were invaluable to this research: HEALPix, Fitsio, Tensorflow, Scikit-Learn, NumPy, SciPy, Nbodykit, Pandas, IPython, Jupyter, GitHub.

%% file: sections/window.tex
%
%
\section{Window Function}\label{app:windowfunction}
The observed density field of targets does not cover the full sky, due to the galactic plane obscuration. This means that the pseudo-power spectrum $\hat{C}_{\ell}$ obtained by the direct Spherical Harmonic Transforms of a partial sky map (eq. \ref{eq:pusedocell}), differs from the full-sky angular spectrum $C_{\ell}$. However, their ensemble average is related by \citep{hivonmaster2002ApJ...567....2H, poker2011A&A...535A..90P} 

\begin{equation}
    <\hat{C}_{\ell}> = \sum_{\ell^{\prime}} M_{\ell \ell^{\prime}}<C_{\ell^{\prime}}>,
\end{equation}
where $M_{\ell \ell^{\prime}}$ represents the mode-mode coupling from the partial sky coverage. This is known as the Window Function effect and a proper assessment of this effect is crucial for a robust measurement of the large-scale clustering of galaxies. We follow a similar approach to that of \citep{szapudi2001ApJ...548L.115S, chon2004MNRAS.350..914C} to model the window function effect on the theoretical power spectrum $C_{\ell}$ rather than correcting the measured pseudo-power spectrum from data. First, we compute the two-point correlation function of the window,

\begin{equation}
    RR(\theta) = \sum_{i,j>1} f_{\rm pix, i} f_{\rm pix, j} \Theta_{ij}(\theta),
\end{equation}
where $\Theta_{ij}(\theta)$ is one when the pixels i and j are separated by an angle between $\theta$ and $\theta + \Delta\theta$, and zero otherwise. Next, we normalize the RR by $\sin(\theta)\Delta\theta$ to account for the area and total number of pairs such that $RR(\theta=0)=1$. We fit a polynomial on RR to smooth out the wiggles raised by noise. Then, we multiply the theoretical correlation function $\omega(\theta)$ by the window paircount,
\begin{align}
    \omega^{WC}(\theta) &= \omega(\theta)~RR(\theta),
\end{align}
and finally use the Gaussian-Quadrature algorithm to transform the window convolved theory correlation function $\omega^{WC}$ to $C^{WC}_{\ell}$,
\begin{equation}
    C^{WC}_{\ell} = 2\pi \int \omega^{WC}(\theta)P_{\ell}(\theta)d\theta.
\end{equation}

\begin{figure}
    \centering
    \includegraphics[width=0.45\textwidth]{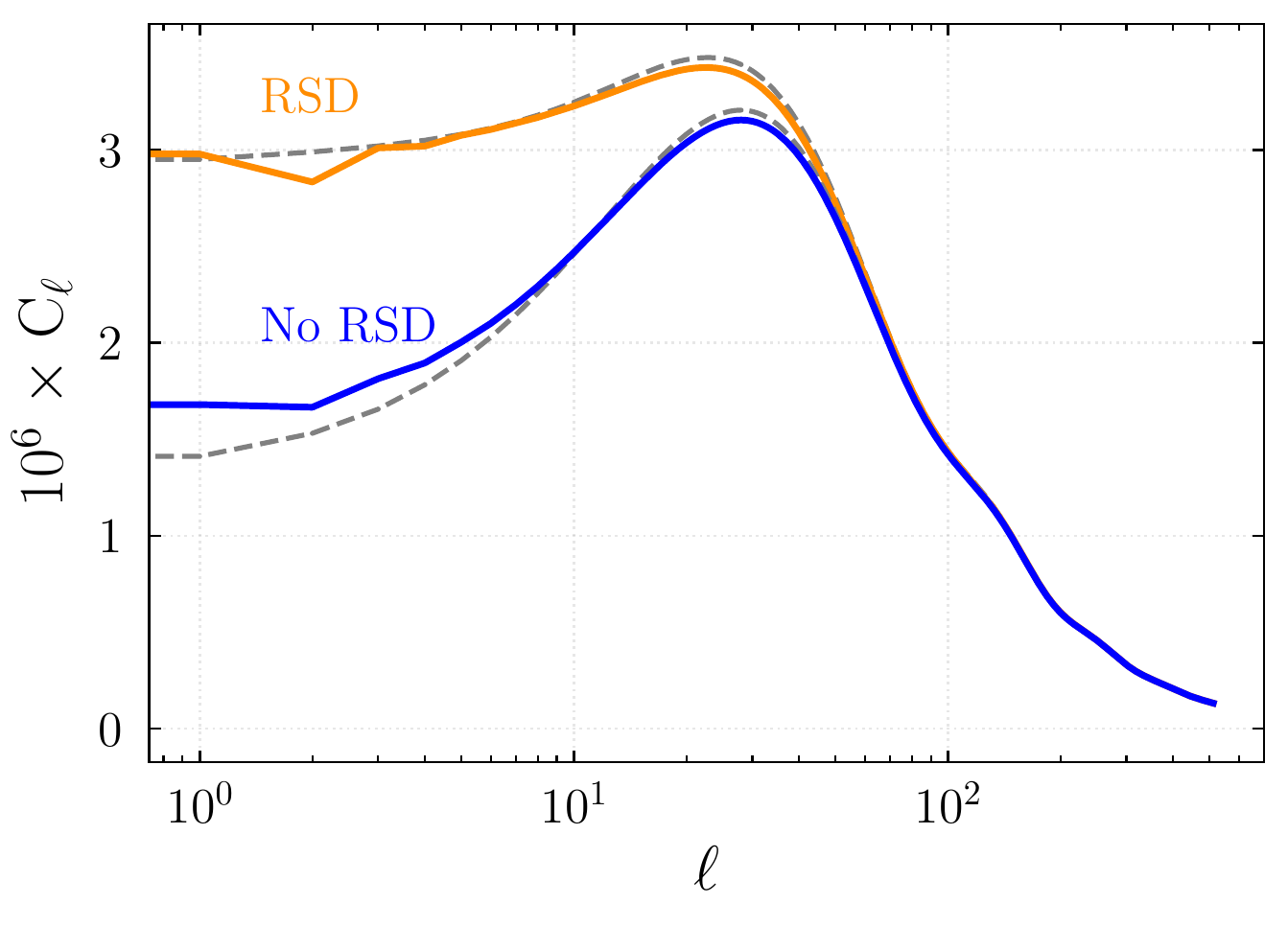}
    \caption{Window corrected theory C$_{\ell}$ for two different models with and without Redshift Space Distortions respectively in orange and blue. The dashed curves show the theoretical models before window convolution. The effect of the window is around 5\% in redshift and 20\% in real space. The theory with redshift space distortions uses $galaxy\;bias = 2$ and the surface density $n(z)$ of NGC eBOSS ELG ~\citep[Tab. 4 of][]{Raichoor2017MNRAS.471.3955R} and assuming the fiducial cosmology of \citet{ashley2012MNRAS,2012ApJ...761...14H}.}
    \label{fig:Cellwindowratio}
\end{figure}

Fig. \ref{fig:Cellwindowratio} shows the DECaLS window effect on two theoretical models of $C_{\ell}$ with and without redshift space distortions. The window effect for the model without Redshift Space Distortions (RSD) is around 20\% but that for the model with RSD is less than 5\% due to the flat power spectrum at the low ell limit. We find a consistent pattern in the mocks; the window effect on the clustering of the mocks is between 5-15\%.